\documentstyle[preprint,aps,prb]{revtex}

\author{Sergey V. Kuplevakhsky\\
Department of Physics, Kharkov State University,\\
310077 Kharkov, Ukraine\\
e-mail: Sergey.V.Kuplevakhsky@univer.kharkov.ua\\
and Institute of Electrical Engineering,\\
SAS, 842 39 Bratislava, Slovak Republic}
\title{Microscopic theory of weakly coupled superconducting multilayers in an external magnetic field
}

\begin{document}

\maketitle
\begin{abstract}
We present the first fully microscopic, self-consistent, and self-contained
theory of superconducting weakly coupled periodic multilayers with tunnel
barriers in the presence of externally applied parallel magnetic fields, in
the local Ginzburg-Landau regime. We solve a nontrivial mathematical problem
of a microscopic derivation and exact minimization of the free-energy
functional. In the thin-layer limit that corresponds to the domain of
validity of the phenomenological Lawrence-Doniach model, our physical
results strikingly contrast with those of our predecessors. In particular,
we completely revise previous calculations of the lower critical field and
refute the concept of a triangular Josephson vortex lattice. We show that
Josephson vortices penetrate into all the barriers simultaneously and form
peculiar structures that we term ''vortex planes''. We calculate the
superheating field of the Meissner state and predict hysteresis in the
magnetization. In the vortex state, the magnetization exhibits distinctive
oscillatory behavior and jumps due to successive penetration of the vortex
planes. We prove that the vortex-plane penetration and pinning by the edges
of the sample cause the Fraunhofer pattern of the critical Josephson
current. We calculate the critical temperature and the upper critical field
of infinite (along the layers ) multilayers. For finite multilayers, we
predict a series of first-order phase transitions to the normal state and
oscillations of the critical temperature versus the applied field. Finally,
we discuss some theoretical and experimental implications of the obtained
results.

\bigskip\

\end{abstract}

\section{INTRODUCTION}

In this paper, we present the first fully microscopic, self-consistent, and
self-contained theory of superconducting weakly coupled periodic multilayers
(superlattices) of the S/I-type (S for a superconductor, I for an insulator
or a semiconductor) in the presence of externally applied parallel magnetic
fields, in the local Ginzburg-Landau\cite{GL} (GL) regime (i. e.,
temperatures are close to the bulk critical temperature $T_{c0}$, all
characteristic dimensions of the S-layers are much larger than the BCS\cite
{BCS} coherence length $\xi _0.$)

Our approach introduces radically new concepts both into the construction of
the theory and understanding of the underlying physics.

In Section II, we derive a microscopic free-energy functional that describes
a smooth transition from the single-junction case to the thin-layer limit,
when the S-layer thickness $a$ is small compared to all other relevant
length scales. [By contrast, previous treatment was based predominantly on
the phenomenological Lawrence-Doniach\cite{LD} (LD) model, applicable only
in the thin-layer limit.] Our analysis of implications of gauge invariance
reveals novel facets of the theory that have been completely overlooked up
to now. In particular, we establish for the first time the existence of
fundamental constraint relations coupling the phases of the order parameter
to the vector potential and equalizing the phase differences at neighboring
barriers. Mathematically, these constraints complement the usual
Euler-Lagrange equations and make the free energy a minimum. Physically,
they state that the average intralayer currents are always equal to zero and
the local magnetic field has, in general, the periodicity of the multilayer.
(The latter has recently received strong experimental support in polarized
neutron reflectivity measurements on artificial Nb/Si multilayers.\cite
{YFOF98}) As a result of this investigation, we obtain a closed,
self-consistent set of mean-field equations. These equations have a
particularly simple form in the thin-layer limit: Remarkably, the phase
differences (the same at all the barriers) obey a Ferrell-Prange-type\cite
{FP63} equation with a single length scale, the Josephson penetration depth $%
\lambda _J=(8\pi epj_0)^{-1/2}$ ($p$ is the period, $j_0$ is the critical
Josephson current). (This should be contrasted with a mathematically
ill-defined infinite set of differential equations containing two length
scales, proposed without appropriate justification in the literature.\cite
{BC91,BCG92}) Furthermore, due to the absence of screening by the intralayer
currents in the thin-layer limit, the local magnetic field proves to be
independent of the coordinate normal to the layers.

In Section III, we obtain exact analytical solutions to the equations of the
thin-layer limit. This limit corresponds to the domain of validity of the
phenomenological LD model, which allows us to draw a comparison with the
results of our predecessors. Unfortunately, for the lack of a closed,
self-consistent set of mean-field equations minimizing the free energy, some
of the previous studies have rather spurious character. Thus, the
above-mentioned fundamental constraints of the theory automatically rule out
any possibility of suggested single Josephson vortex penetration\cite
{B73,CC90} as well as the occurrence\cite{BC91} of a triangular Josephson
vortex lattice. Another example: in Ref. (\cite{BCG92}) concerned with the
Josephson effect, inability to carry out self-consistent calculations misled
the authors into an absolutely incorrect conclusion that the Fraunhofer
pattern of the total critical current occurred in the absence of Josephson
vortices. Here we undertake a complete revision of these issues. However,
our consideration contains as a limiting case a particular exact solution
for the vortex state obtained in the framework of the LD model by Theodorakis%
\cite{Th90} and fully supports phenomenological calculations\cite{DE78,SS93}
of the upper critical field $H_{c2\infty }$ in infinite (along the layers)
multilayers.

Among new physical results of Sec. III, there are the following. We provide
the first comprehensive description of the Meissner state in semi-infinite
(along the layers) multilayers and show that at the field $H_s=(ep\lambda
_J)^{-1}$ (the superheating field) the Meissner phase becomes unstable with
regard to Josephson vortex penetration. We predict simultaneous and coherent
penetration into all the barriers. (This prediction has been confirmed
experimentally.\cite{YFOF98}) We show that in the absence of screening by
the intralayer currents (see above) the ''tails'' of Josephson vortices
overlap in the layering direction forming peculiar structures, ''vortex
planes''. The lower critical field at which the formation of a single vortex
plane becomes energetically favorable in an infinite multilayer is found to
be $H_{c1\infty }=2(\pi ep\lambda _J)^{-1}$. For the lower critical field in
a finite multilayer with $W\ll \lambda _J$ ($W$ is the S-layer length) we
obtain $H_{c1W}=\pi /epW$, which corresponds to the first minimum of the
Fraunhofer pattern. We prove that the Fraunhofer oscillations occur due to
successive penetration of the vortex planes and their pinning by the edges
of the sample. We show that vortex-plane penetration leads also to jumps of
the magnetization. (Such features have been already observed.\cite{YFOF98})
For a certain field range, we predict a small paramagnetic effect. We
calculate the critical temperature and the upper critical field of an
infinite multilayer. The obtained implicit dependence $H_{c2\infty }(T)$
exhibits the well-known ''3D-2D crossover'' and is free from the unphysical
''low-temperature'' divergence\cite{T96} of the LD model. In addition, we
predict novel size-effects in finite multilayers: a series of first-order
phase transitions to the normal state and oscillations of the critical
temperature versus the applied field.

In Section IV, we discuss some theoretical and experimental implications of
the obtained results. In Appendix A, we write down a few mathematical
formulas related to the application of Mathieu functions in Sec. III.

\section{BASIC EQUATIONS OF THE THEORY}

\subsection{A derivation and exact minimization of the microscopic
free-energy functional}

Our starting point is a microscopic second-quantized BCS-type Hamiltonian of
the form\cite{KN97,KNG97}

$$
H={\int\limits_R^{}}d^3{\bf r}\psi _\alpha ^{+}({\bf r})\left[ -\frac
1{2m}\left[ {\bf \vec \nabla }-ie{\bf \tilde A(r)-}ie{\bf A}_{ext}({\bf r}
)\right] ^2-E_F\right] \psi _\alpha ({\bf r})
$$

$$
-\frac{\left| g\right| }2{\int\limits_{R_s}^{}}d^3{\bf r}g({\bf r})\psi
_\alpha ^{+}({\bf r})\psi _{-\alpha }^{+}({\bf r})\psi _{-\alpha }({\bf r}
)\psi _\alpha ({\bf r})+{\int\limits_{R_s}^{}}d^3{\bf r}\psi _\alpha ^{+}(
{\bf r})V_{imp}({\bf r})\psi _\alpha ({\bf r})
$$
\begin{equation}
\label{1.1}+U_0{\int\limits_{R_b}^{} }d^3{\bf r}\psi _\alpha ^{+}({\bf r}%
)\psi _\alpha ({\bf r})+\frac 1{8\pi }{\int\limits_{R}^{} }d^3{\bf r\tilde h}%
^2{\bf (r)},
\end{equation}
$$
R_s={\cup _{n=-\infty }^{+\infty }}R_{s_n},R_b=%
{\cup _{n=-\infty }^{+\infty }}R_{b_n},R=R_s\cup R_b,
$$
$$
R_{s_n}=[-a/2+np\leq x\leq a/2+np]\times [L_{y1}\leq y\leq L_{y2}]\times
(-\infty <z<+\infty ),
$$
$$
R_{b_n}=[a/2+(n-1)p\leq x\leq -a/2+np]\times [L_{y1}\leq y\leq L_{y2}]\times
(-\infty <z<+\infty ),
$$
\begin{equation}
\label{1.2}{\bf \tilde h(r)}={\bf \vec \nabla \times \tilde A(r)},\text{ }%
{\bf H}={\bf \vec \nabla \times A}_{ext}{\bf (r)}\equiv (0,0,H){\bf ,}
\end{equation}
$$
{\bf \vec \nabla }=(\nabla _x,\nabla _y,\nabla _z)\equiv (\frac \partial
{\partial x},\frac \partial {\partial y},\frac \partial {\partial z}).
$$

Here $\hbar =c=1$, $E_F=k_F^2/2m$ is the Fermi energy (with $k_F$ being the
Fermi momentum), $R_s$ and $R_b$ correspond respectively to the
superconducting and barrier regions (with $a$ being the S-layer thickness, $%
b $ the barrier thickness and $p=a+b$ the period, the $x$ axis being normal
to the barrier interfaces), $\psi _\alpha ({\bf r})$ is the electron field
operator for spin $\alpha $ (a summation over repeated spin indices is
implied), $g<0$ is the BCS coupling constant, $V_{imp}({\bf r})$ is the
nonmagnetic impurity potential, $U_0>0$ is the repulsive barrier potential,
{\bf A}$_{ext}$ and {\bf \~A} are the external (classical) and induced
(operator) vector potentials.\cite{BD} The system is taken to be infinite in
the direction of the $x$ and $z$ axes, while no restrictions on the linear
dimensions along the $y$ axis is so far imposed. The external magnetic field
${\bf H}$ is directed along the $z$ axis (see Fig.~1).

Using field-theoretical methods of Ref. (\cite{KN97}), we can derive from (%
\ref{1.1}) a microscopic free-energy functional of the system $\Omega \left[
\Delta _n,\Delta _n^{*},{\bf A};H\right] $, where $\Delta _n$ and ${\bf A}$
are classical variables: $\Delta _n$ is the pair potential (order parameter)
of the $n$th S-layer, and ${\bf A=\tilde A+A}_{ext}$ is the total vector
potential, ${\bf h(r)}={\bf \vec \nabla \times A(r)}$ being the
corresponding local magnetic field. For external fields satisfying the
quasiclassical condition $H\ll k_F/e\xi _0$, in the GL regime
\begin{equation}
\label{1.3}\tau \equiv \frac{T_{c0}-T}{T_{c0}}\ll 1,
\end{equation}
\begin{equation}
\label{1.4}\xi _0\ll a,\text{ }W\equiv L_{y2}-L_{y1},
\end{equation}
where $T_{c0}$ is the bulk critical temperature, $\xi _0=v_0/2\pi T_{c0}$ is
the BCS coherence length ($v_0=k_F/m$), this functional takes on the form

$$
\Omega \left[ f_n,\phi _n,A_x,A_y;H\right] =\frac{H_c^2(T)}{4\pi }W_z%
{\int\limits_{L_{y1}}^{L_{y2}}}dy\left[
{\sum _{n=-\infty }^{+\infty }}{\int\limits_{-a/2+np}^{a/2+np}}
dx\left[ -f_n^2(x,y)+\frac 12f_n^4(x,y)\right. \right.
$$

$$
+\zeta ^2(T){\sum _{i=x,y}}\left[ \left[ \nabla _if_n(x,y)\right]
^2+\left[ \nabla _i\phi _n(x,y)-2eA_i(x,y)\right] ^2f_n^2(x,y)\right]
$$

$$
+\frac{\alpha \zeta ^2(T)}{2a\xi _0}\left[ f_{n-1}^2\left(
a/2+(n-1)p,y\right) +f_n^2\left( -a/2+np,y\right) \right.
$$

$$
\left. -2f_n\left( -a/2+np,y\right) f_{n-1}\left( a/2+(n-1)p,y\right) \cos
\Phi _{n,n-1}(y)\right]
$$
\begin{equation}
\label{1.5}\left. +4e^2\zeta ^2(T)\lambda ^2(T){\int\limits_{L_{x1}}^
{L_{x2}}}dx\left[ h(x,y)-H\right] ^2\right] ,
\end{equation}
$$
\Phi _{n,n-1}(y)=\phi _n\left( -a/2+np,y\right) -\phi _{n-1}\left(
a/2+(n-1)p,y\right) -2e{\int\limits_{a/2+(n-1)p}^{-a/2+np}}
dxA_x(x,y),
$$
$$
\alpha =\frac{3\pi ^2}{7\zeta (3)\chi \left( \xi _0/l\right) }
{\int\limits_0^1}dttD(t),
$$
$$
D(t)=\frac{16E_Ft^2\left( U_0-E_Ft^2\right) }{U_0^2}\exp \left[ -2b\sqrt{%
2m\left( U_0-E_Ft^2\right) }\right] ,
$$
$$
\chi \left( \xi _0/l\right) =\frac 8{7\zeta (3)}
{\sum _{n=0}^{+\infty }}\left( 2n+1\right) ^{-2}\left( 2n+1+\xi
_0/l\right) ^{-1},
$$
$$
H_c^2(T)=4\pi N(0)\Delta _\infty ^2(T)\tau ,
$$
\begin{equation}
\label{1.6}h(x,y)=\frac{\partial A_y(x,y)}{\partial x}-\frac{\partial
A_x(x,y)}{\partial y}.
\end{equation}
In Eq. (\ref{1.5}), we have introduced the reduced modulus $0\leq f_n\leq 1$
and the phase $\phi _n$ of the pair potential in the $n$th S-layer via the
relation $\Delta _n=\Delta _\infty f_n\exp (i\phi _n)$, where $\Delta
_\infty (T)=\sqrt{\frac{8\pi ^2T_{c0}^2\tau }{7\zeta (3)}}$ is the bulk gap,
$\zeta (m)$ is the Riemann zeta function.\cite{AS} The rest of notation are
as follows: $W_z=L_{z2}-L_{z1}$ is the length of the system in the $z$
direction, $D(t)$ is the tunneling probability of an insulating barrier
between two successive S-layers [$D(1)\ll 1$], $\chi \left( \xi _0/l\right) $
is the impurity factor\cite{AGD63} ($l$ is the electron mean free path), $%
\zeta (T)=\xi _0\sqrt{\frac{7\zeta (3)\chi \left( \xi _0/l\right) }{12\tau }}
$ is the GL coherence length, $\lambda (T)=\frac{\sqrt{3}\left[ \pi \chi
\left( \xi _0/l\right) \xi _0N(0)\tau \right] ^{-1/2}}{8\pi eT_{c0}}$ is the
GL penetration depth, $N(0)=mk_F/2\pi ^2$ is the one-spin density of states
at the Fermi level, and $H_c(T)$ is the bulk thermodynamic critical field
near $T_{c0}$.\cite{AGD63} The term proportional to $\alpha \ll 1$
determines the interlayer Josephson coupling. Equation (\ref{1.6}) is merely
the Maxwell equation for the local magnetic field ${\bf h}=(0,0,h)$.

The microscopic free-energy functional (\ref{1.5}) covers all well-known
limiting cases. In the limit $\alpha =0$ (no Josephson interlayer coupling),
equation (\ref{1.5}) reduces to a sum of free-energy functionals of
independent S-layers. Making a shift of the coordinate system $x\rightarrow
x-a/2-b/2$ and taking the limit $a\rightarrow \infty $, one gets the case of
a single SIS junction. Shifting $x\rightarrow x-a/2$ and taking $%
a\rightarrow \infty $, $b\rightarrow \infty $, we recover the limit of a
semi-infinite superconductor in contact with vacuum.

Our task now is to establish mean-field equations of the theory, which is
mathematically equivalent to the problem of minimization of (\ref{1.5}) with
respect to $f_n$, $\phi _n$, and ${\bf A}=(A_x,A_y,0)$. This problem should
be approached with certain caution, because Euler-Lagrange equations for $%
\phi _n$, and $A_x,$ $A_y$ are not independent.

Indeed, the functional (\ref{1.5}) is invariant under the general gauge
transformation%
$$
\phi _n(x,y)\rightarrow \phi _n(x,y)+\eta (x,y),
$$
\begin{equation}
\label{1.7}A_i(x,y)\rightarrow A_i(x,y)+\frac 1{2e}\nabla _i\eta (x,y),
\end{equation}
where $\eta (x,y)$ is an arbitrary gauge function, defined in the whole
region $R$. As a result, the variational derivatives with respect to $\phi
_n $, and $A_x,$ $A_y$ are related through fundamental functional identities
\begin{equation}
\label{1.8}\frac{\delta \Omega }{\delta \phi _n(x,y)}\equiv \frac 1{2e}%
{\sum _{i=x,y}}\nabla _i\frac{\delta \Omega }{\delta A_i(x,y)}.
\end{equation}
The occurrence of such identities is typical of gauge theories.\cite{GT86}
Moreover, identities relating variational derivatives appear already in some
problems of classical variational calculus with degenerate (i. e., invariant
under symmetry transformations) functionals.\cite{Sch67} As in degenerate
theories the number of variables exceeds the number of independent
Euler-Lagrange equations, complementary relations should be normally imposed
to eliminate irrelevant degrees of freedom and close the system
mathematically. Whereas in bulk superconductors and single junctions the
elimination of unphysical degrees of freedom amounts merely to an
appropriate choice of gauge, in periodic weakly coupled structures this
problem has additional implications. Namely, in the presence of the
Josephson interlayer coupling phase differences $\Phi _{n,n-1}$ and $\Phi
_{n+1,n}$ at two successive barriers are in themselves not independent,
which means, mathematically, that we are dealing with a variational problem
with constraints. Unfortunately, this fundamental feature was completely
overlooked in previous literature.

The variations with respect to $f_n$ are independent and can be taken first.
Varying under the assumption of arbitrary $\delta f_n$ at the boundaries, we
obtain

\begin{equation}
\label{1.9}\left[ 1+\zeta ^2(T){\sum _{i=x,y} }\left[ \nabla
_i^2-\left[ \nabla _i\phi _n(x,y)-2eA_i(x,y)\right] ^2\right] \right]
f_n(x,y)-f_n^3(x,y)=0,
\end{equation}
$$
(x,y)\in R_{s_n};
$$
\begin{equation}
\label{1.10}\frac{\partial f_n}{\partial y}(x,L_{y1})=\frac{\partial f_n}{%
\partial y}(x,L_{y2})=0;
\end{equation}
\begin{equation}
\label{1.11}\frac{\partial f_n}{\partial x}\left( -a/2+np,y\right) =\frac
\alpha {2\xi _0}\left[ f_n\left( -a/2+np,y\right) -f_{n-1}\left(
a/2+(n-1)p,y\right) \cos \Phi _{n,n-1}(y)\right] ,
\end{equation}
\begin{equation}
\label{1.12}\frac{\partial f_n}{\partial x}\left( a/2+np,y\right) =-\frac
\alpha {2\xi _0}\left[ f_n\left( a/2+np,y\right) -f_{n+1}\left(
-a/2+(n+1)p,y\right) \cos \Phi _{n+1,n}(y)\right] .
\end{equation}
Here, equation (\ref{1.9}) is the usual GL equation for the bulk order
parameter. Relations (\ref{1.10}) are the usual GL boundary conditions at
the superconductor/vacuum interfaces. Boundary conditions (\ref{1.11}), (\ref
{1.12}), describing the suppression of the order parameter due to the
Josephson currents at the superconductor/insulator interfaces, are of the
type first derived by de Gennes\cite{dG66} for a single junction.

By contrast to $f_n$, the variables $A_x,$ $A_y$ are defined in the whole
region $R$. For these variables, continuity up to the second-order partial
derivatives at the superconductor/insulator interfaces should be assumed.
The corresponding Euler-Lagrange equations are
\begin{equation}
\label{1.13}-\frac{\partial h(x,y)}{\partial x}=\frac 1{2e}\frac{f_n^2(x,y)}{%
\lambda ^2(T)}\left[ \frac{\partial \phi _n(x,y)}{\partial y}%
-2eA_y(x,y)\right] \equiv 4\pi j_{ny}(x,y),
\end{equation}
\begin{equation}
\label{1.14}\frac{\partial h(x,y)}{\partial y}=\frac 1{2e}\frac{f_n^2(x,y)}{%
\lambda ^2(T)}\left[ \frac{\partial \phi _n(x,y)}{\partial x}%
-2eA_x(x,y)\right] \equiv 4\pi j_{nx}(x,y),\text{ }(x,y)\in R_{s_n};
\end{equation}
\begin{equation}
\label{1.15}\frac{\partial h(x,y)}{\partial y}=4\pi j_0f_n\left(
-a/2+np,y\right) f_{n-1}\left( a/2+(n-1)p,y\right) \sin \Phi
_{n,n-1}(y)\equiv 4\pi j_{n,n-1}(y),
\end{equation}
\begin{equation}
\label{1.16}j_0=\frac{7\zeta (3)\alpha \chi \left( \xi _0/l\right) }%
6eN(0)\xi _0\Delta _\infty ^2(T),
\end{equation}
\begin{equation}
\label{1.17}\frac{\partial h(x,y)}{\partial x}=0,\text{ }(x,y)\in R_{b_n}.
\end{equation}
Here, equations (\ref{1.13}), (\ref{1.14}) are the Maxwell equations in the
S-layers, with ${\bf j}_n$ being the intralayer supercurrent densities.
Equations (\ref{1.15}), (\ref{1.17}) are the Maxwell equations in the
barrier regions, with $j_{n,n-1}$ being the Josephson current density
between the $n$th and the $(n-1)$th layers. Relation (\ref{1.16}) is the
definition of the Josephson critical current density in a single SIS
junction.\cite{KN97}

Equations (\ref{1.13})-(\ref{1.17}) should be complemented by boundary
conditions at the outer interfaces $y=L_{y1,}L_{y2}$. [When deriving these
equations, we have only assumed $\delta A_x(x,L_{y1})=\delta A_x(x,L_{y2})=0$%
.] As we do not consider here externally applied currents in the $y$
direction, the first set of boundary conditions follows from the requirement
$\left[ j_{ny}\right] _{y=L_{y1,}L_{y2}}=0$:
\begin{equation}
\label{1.18}\left[ \frac{\partial \phi _n(x,y)}{\partial y}%
-2eA_y(x,y)\right] _{y=L_{y1},L_{y2}}=0.
\end{equation}
Applied to Eq. (\ref{1.13}), these boundary conditions show that the local
magnetic field at the outer interfaces is independent of the coordinate $x$:
$h(x,L_{y1})=h(L_{y1})$, $h(x,L_{y2})=h(L_{y2})$. The boundary conditions
imposed on $h$ should be compatible with Ampere's law $h(L_{y2})-h(L_{y1})=4%
\pi I$ obtained by integration of Eqs. (\ref{1.13}), (\ref{1.15}) over $y$,
where
\begin{equation}
\label{1.19}I\equiv {\int\limits_{L_{y1}}^{L_{y2}} }
dyj_{nx}(x,y)={\int\limits_{L_{y1}}^{L_{y2}}}dyj_{n,n-1}(y)
\end{equation}
is the total current in the $x$ direction. Throughout this paper, depending
on a physical situation under consideration, we will employ three types of
boundary conditions on $h$:
\begin{equation}
\label{1.20}\left\{
\begin{array}{c}
h(L_{y1})=h(L_{y2})=H,
\text{ (i);} \\ h(L_{y1})=H-2\pi I,\quad h(L_{y2})=H+2\pi I,
\text{ (ii);} \\ h(L_{y1})=H,\quad h(L_{y2})=H+4\pi I,\text{ (iii).}
\end{array}
\right.
\end{equation}

As usual, the Maxwell equations (\ref{1.13}), (\ref{1.14}) yield the
current-continuity equations inside the S-layers:
\begin{equation}
\label{1.21}{\sum _{i=x,y}}\nabla _i\left[ f_n^2(x,y)\left[ \nabla
_i\phi _n(x,y)-2eA_i(x,y)\right] \right] =0.
\end{equation}
The conservation of Josephson interlayer current is readily verified from (%
\ref{1.15}). Using Eqs. (\ref{1.14}), (\ref{1.15}) and assumed continuity of
$\partial h/\partial y$, we arrive at the boundary conditions
\begin{equation}
\label{1.22}\left[ \left[ \frac{\partial \phi _n(x,y)}{\partial x}%
-2eA_x(x,y)\right] f_n(x,y)\right] _{x=-a/2+np}=\frac \alpha {2\xi
_0}f_{n-1}\left( a/2+(n-1)p,y\right) \sin \Phi _{n,n-1}(y),
\end{equation}
\begin{equation}
\label{1.23}\left[ \left[ \frac{\partial \phi _n(x,y)}{\partial x}%
-2eA_x(x,y)\right] f_n(x,y)\right] _{x=a/2+np}=\frac \alpha {2\xi
_0}f_{n+1}\left( -a/2+(n+1)p,y\right) \sin \Phi _{n+1,n}(y),
\end{equation}
reflecting the continuity of the $x$ component of the supercurrent at the
internal interfaces $x=\pm a/2+np$.

Integrating Eqs. (\ref{1.9}), (\ref{1.21}) over $x$ and applying boundary
conditions (\ref{1.11}), (\ref{1.12}) and (\ref{1.22}), (\ref{1.23}),
respectively, we obtain very useful integro-differential representations%
$$
\overline{f_n(x,y)}-\overline{f_n^3(x,y)}-\zeta ^2(T){\sum _{i=x,y}}
\overline{\left[ \nabla _i\phi _n(x,y)-2eA_i(x,y)\right] ^2f_n(x,y)}+\zeta
^2(T)\frac{\partial ^2\overline{f_n(x,y)}}{\partial y^2}
$$
$$
=\frac{\alpha \zeta ^2(T)}{2a\xi _0}\left[ f_n\left( -a/2+np,y\right)
+f_n\left( a/2+np,y\right) \right.
$$
\begin{equation}
\label{1.24}\left. -f_{n+1}\left( -a/2+(n+1)p,y\right) \cos \Phi
_{n+1,n}(y)-f_{n-1}\left( a/2+(n-1)p,y\right) \cos \Phi _{n,n-1}(y)\right] ,
\end{equation}
$$
\frac{\partial \overline{\left[ f_n^2(x,y)\left[ \frac{\partial \phi _n(x,y)
}{\partial y}-2eA_y(x,y)\right] \right] }}{\partial y}
$$
$$
=\frac \alpha {2a\xi _0}\left[ f_n\left( -a/2+np,y\right) f_{n-1}\left(
a/2+(n-1)p,y\right) \sin \Phi _{n,n-1}(y)\right.
$$
\begin{equation}
\label{1.25}\left. -f_n\left( a/2+np,y\right) f_{n+1}\left(
-a/2+(n+1)p,y\right) \sin \Phi _{n+1,n}(y)\right] ,
\end{equation}
where $\overline{\left( x,\ldots \right) }\equiv \frac 1a
{\int\limits_{-a/2+np}^{a/2+np}}\left( x,\ldots \right) dx$ denotes averaging
over the interval $-a/2+np\leq x\leq a/2+np$.

By summing Eqs. (\ref{1.25}) over the layer index $n$, integrating and
applying boundary conditions (\ref{1.18}), we obtain the integral
\begin{equation}
\label{1.27}{\sum _{n=-\infty }^{+\infty }}\overline{%
f_n^2(x,y)\left[ \frac{\partial \phi _n(x,y)}{\partial y}-2eA_y(x,y)\right] }%
=0,
\end{equation}
which is, physically, the conservation law for the total supercurrent in the
$y$ direction. Mathematically, equation (\ref{1.27}) has the form of a
constraint relation between variables $\partial $$\phi _n/\partial y$ and $%
A_y$.\cite{GT86,L62} To find the rest of constraints of the theory, closing
the system of equations, we must minimize the functional (\ref{1.5}) with
respect to $\phi _n$ and $\partial $$\phi _n/\partial y.$

By virtue of fundamental identities (\ref{1.8}), a naive variation of (\ref
{1.5}) with respect to $\phi _n$ (with arbitrary $\delta \phi _n$ at the
boundaries) does not yield new equations. Indeed, the corresponding
Euler-Lagrange equation reduces to the conservation law (\ref{1.21}), while
surface variations merely reproduce boundary conditions (\ref{1.18}), and (%
\ref{1.22}), (\ref{1.23}). Considering variations of the type $\phi
_n(x,y)\rightarrow \phi _n(x,y)+\epsilon \psi _n(y)$, $\partial $$\phi
_n(x,y)/\partial y\rightarrow \partial \phi _n(x,y)/\partial y+\epsilon
\partial \psi _n(y)/\partial y$, where $\epsilon $ is a small parameter and $%
\psi _n(y)$ are arbitrary functions of $y$, we arrive at Eqs. (\ref{1.25}).
To obtain genuinely new equations, minimizing (\ref{1.5}) with respect to $%
\phi _n$ and $\partial $$\phi _n/\partial y$, we must enlarge the class of
allowed variations.

A mathematically rigorous approach to this problem is as follows. While
varying $\phi _n(x,y)\rightarrow \phi _n(x,y)+\delta \phi _n(y)$, $\partial $%
$\phi _n(x,y)/\partial y\rightarrow \partial \phi _n(x,y)/\partial
y+\partial \delta \phi _n(y)/\partial y$, with $\delta \phi _n(y)$ being
small arbitrary functions of $y$, instead of integrating by parts, we impose
additional constraints
\begin{equation}
\label{1.28}\overline{f_n^2(x,y)\left[ \frac{\partial \phi _n(x,y)}{\partial
y}-2eA_y(x,y)\right] }=0,
\end{equation}
compatible with boundary conditions (\ref{1.18}) and constraint relation (%
\ref{1.27}). The requirement of compatibility with the current-conservation
law (\ref{1.25}) automatically yields another set of constraints%
$$
f_n\left( -a/2+np,y\right) f_{n-1}\left( a/2+(n-1)p,y\right) \sin \Phi
_{n,n-1}(y)
$$
\begin{equation}
\label{1.29}=f_n\left( a/2+np,y\right) f_{n+1}\left( -a/2+(n+1)p,y\right)
\sin \Phi _{n+1,n}(y).
\end{equation}
The above procedure is formally equivalent to minimization of (\ref{1.5})
with respect to independent variations of $\phi _n$ and $\partial $$\phi
_n/\partial \dot y$. As this class of variations of $\phi _n$ and $\partial $%
$\phi _n/\partial \dot y$ is larger than that employed in deriving (\ref
{1.25}), we can argue that Eqs. (\ref{1.28}) and (\ref{1.29}) provide the
sought necessary conditions for the true minimum of the free-energy
functional (\ref{1.5}).

The physical meaning of Eqs. (\ref{1.28}) and (\ref{1.29}) is quite
transparent. Constraints (\ref{1.28}) minimize the kinetic-energy term in (%
\ref{1.5}) with respect to variations $\partial $$\phi _n(x,y)/\partial
y\rightarrow \partial \phi _n(x,y)/\partial y+\delta \psi _n(y)$, where $%
\delta \psi _n(y)$ are small arbitrary functions of $y$. They show that the
average intralayer currents in the $y$ direction are always equal to zero,
and, as a result,
\begin{equation}
\label{1.30}h\left( -a/2+np,y\right) =h\left( a/2+np,y\right) .
\end{equation}
[See Eq. (\ref{1.13})]. These constraints appear already in the case of
decoupled S-layers. By contrast, constraints (\ref{1.29}) are uniquely
imposed by the Josephson interlayer coupling. Their function is to make the
Josephson energy stationary with respect to variations $\phi
_n(x,y)\rightarrow \phi _n(x,y)+\delta \phi _n(y)$ and to assure the
conservation of the total Josephson current $I$ in neighboring barriers [see
Eqs. (\ref{1.15}), (\ref{1.19})].

As no other conditions are imposed on the variables, we can satisfy (\ref
{1.29}) by choosing
\begin{equation}
\label{1.31}f_n(x,y)=f_{n-1}(x-p,y)\equiv f(x,y),\text{ }f(x+np,y)=f(x,y),
\end{equation}
\begin{equation}
\label{1.32}\Phi _{n+1,n}(y)=\Phi _{n,n-1}(y)\equiv \Phi (y).
\end{equation}
These relations finalize the determination of a closed, complete,
self-consistent system of mean-field equations for a S/I superlattice in the
GL regime.

Constraints (\ref{1.28}), (\ref{1.29}) and their corollaries (\ref{1.30})-(%
\ref{1.32}) belong to key results of this paper. Derived by means of a
rigorous mathematical analysis of the impact of gauge invariance, they are
not restricted to the functional (\ref{1.5}), but should hold for any
superconducting weakly coupled periodic structure. To illustrate their
importance, we point out that Eqs. (\ref{1.29}), (\ref{1.30}), for example,
completely rule out any possibility of single Josephson vortex penetration%
\cite{B73,CC90} and triangular Josephson vortex lattice,\cite{BC91} proposed
without appropriate physical and mathematical justification. On the
contrary, they imply that the distribution of the local magnetic field due
to the Josephson vortices has, in general, the periodicity of the
multilayer, as recently verified experimentally.\cite{YFOF98} It should be
noted, however, that although the role of constraints (\ref{1.29}), (\ref
{1.30}) in minimizing the free energy and closing the system of
Euler-Lagrange equations for $f_n$, $\phi _n$, and ${\bf A}$ has not been
realized until now, relations (\ref{1.31}), (\ref{1.32}) were implicitly
employed in phenomenological calculations of $H_{c2\infty }.$\cite
{LD,DE78,SS93,T96,KLB75} Moreover, relations of the type (\ref{1.28}), (\ref
{1.31}), (\ref{1.32}) were used by Theodorakis\cite{Th90} in his particular
exact solution of the LD model in a parallel field.

The equations of this subsection admit exact solutions in two limiting
situations: the single-junction case, when $a\gg \max \{\zeta (T),\lambda
(T)\}$; the thin-layer limit, when $\xi _0\ll a\ll \min \{\zeta (T),\lambda
(T),\alpha ^{-1}\xi _0,W\}$ ($W\equiv W_y=L_{y2}-L_{y1}$ is the length of
the S-layer in the $y$ direction). The single-junction case is well-known.
The thin-layer limit will be extensively discussed in the next subsection
and in Sec. III.

\subsection{The thin-layer limit}

The mean-field equations of the previous subsection allow remarkable
simplification in the thin-layer limit, when $\xi _0\ll a\ll \min \{\zeta
(T),\lambda (T),\alpha ^{-1}\xi _0,W\}.$

First, we can neglect the $x$-dependence of $f,$ defined by Eq. (\ref{1.31}%
): $f(x,y)\equiv f(y)$. Second, fixing the gauge by the condition
\begin{equation}
\label{1.33}A_x(x,y)\equiv 0,\text{ }A_y(x,y)\equiv A(x,y),
\end{equation}
we can neglect the $x$-dependence of $\phi _n$ as well: $\phi _n(x,y)\equiv
\phi _n(y)$. In the gauge (\ref{1.33}), $\Phi _{n,n-1}(y)=\phi _n(y)-\phi
_{n-1}(y)$, and Eqs. (\ref{1.32}) become
$$
\phi _{n+1}(y)+\phi _{n-1}(y)=2\phi _n(y),\phi _n(y)-\phi _{n-1}(y)=\phi
(y),
$$
with the solution
\begin{equation}
\label{1.34}\phi _n(y)=n\phi (y)+\chi (y),
\end{equation}
where $\phi (y)$ is the coherent phase difference (the same for all the
barriers), and $\chi (y)$ is an arbitrary gauge function, allowed by
particular gauge transformations $\phi _n(y)\rightarrow \phi _n(y)+\chi (y)$%
, $A(x,y)\rightarrow A(x,y)+\frac 1{2e}\frac{\partial \chi (y)}{\partial y}.$
Without any loss of generality, we can set $\chi \equiv 0$.

In view of independence of $f$ and $\phi _n$ from $x$ in the thin-layer
limit, the physical meaning of constraints (\ref{1.28}), (\ref{1.29})
becomes even more obvious. Thus, Eqs. (\ref{1.29}) are now the conditions of
stationarity of the Josephson energy with respect to all allowed variations
of $\phi _n$. Due to Eqs. (\ref{1.28}), the term in (\ref{1.24}) responsible
for the kinetic energy of the intralayer currents becomes%
$$
\zeta ^2(T){\sum _{i=x,y} }\overline{\left[ \nabla _i\phi
_n(x,y)-2eA_i(x,y)\right] ^2f(x,y)}\rightarrow \zeta ^2(T)\overline{\left[
\frac{d\phi _n(y)}{dy}-2eA(x,y)\right] ^2}f(y)
$$
$$
=\zeta ^2(T)\left[ \frac{d\phi _n(y)}{dy}-2e\overline{A(x,y)}\right]
^2f(y)+4e^2\zeta ^2(T)\left[ \overline{A^2(x,y)}-\overline{A(x,y)}^2\right]
f(y)
$$
$$
=4e^2\zeta ^2(T)\left[ \overline{A^2(x,y)}-\overline{A(x,y)}^2\right] f(y),
$$
which shows that conditions (\ref{1.28}) minimize the kinetic energy for a
given configuration of the vector potential $A$.

Concerning the Maxwell equations, the right-hand side of Eq. (\ref{1.13}) is
of order $a^2/\lambda ^2(T)$ and can be discarded. Equation (\ref{1.14}) can
be altogether dropped. Thus, we arrive at a closed set of equations
\begin{equation}
\label{1.35}\frac{\partial ^2A(x,y)}{\partial x^2}=0,\text{ }(x,y)\in R;
\end{equation}
\begin{equation}
\label{1.36}\frac 1{4\pi }\frac{\partial ^2A(x,y)}{\partial y\partial x}%
=j_0f^2(y)\sin \phi (y),\text{ }(x,y)\in R_{b_n};
\end{equation}
\begin{equation}
\label{1.37}n\frac{d\phi (y)}{dy}-2e\overline{A(x,y)}=0,\text{ }(x,y)\in
R_{s_n};
\end{equation}
\begin{equation}
\label{1.38}h(x,y)=\frac{\partial A(x,y)}{\partial x},\text{ }(x,y)\in R.,
\end{equation}
$$
f(y)-f^3(y)-4e^2\zeta ^2(T)\left[ \overline{A^2(x,y)}-\overline{A(x,y)}%
^2\right] f(y)+\zeta ^2(T)\frac{d^2f(y)}{dy^2}
$$
\begin{equation}
\label{1.39}=\frac{\alpha \zeta ^2(T)}{a\xi _0}\left[ 1-\cos \phi (y)\right]
f(y),\text{ }(x,y)\in R_{s_n}.
\end{equation}
These equations, of cause, should be complemented by continuity conditions
on $A$, $\partial $$A/\partial x$ and boundary conditions (\ref{1.10}) and (%
\ref{1.20}).

It is worth noting that an immediate consequence of Eqs. (\ref{1.35}), (\ref
{1.39}) is independence of the local field $h$ from the coordinate $x$ in
the whole region $R$: $h(x,y)=h(y)$, $-\infty <x<+\infty $. This result is
fully compatible with the requirement (\ref{1.30}) and demonstrates that the
intralayer supercurrents in the thin-layer limit are unable to screen out
the magnetic field: The situation is very familiar from the physics of
isolated superconducting films with $a\ll \lambda (T)$.\cite{T96,dG66,A88}

Our next objective is to eliminate the vector potential and obtain a closed
set of equations involving only $f$ and $\phi $. Equations (\ref{1.25}), (%
\ref{1.36}) can be easily solved for $A$ in the $n$th ''elementary cell'' $%
R_n=R_{s_n}\cup R_{b_n}$ (the S-layer plus the adjacent barrier). Applying
the continuity conditions on $A$, $\partial $$A/\partial x$, boundary
conditions (\ref{1.20}) and the constraint relation (\ref{1.37}), we get
\begin{equation}
\label{1.40}A(x,y)=\left[ 4\pi j_0{\int\limits_{L_{y1}}^{y}}
duf^2(u)\sin \phi (u)+H_1\right] \left( x-np\right) +\frac n{2e}\frac{d\phi
(y)}{dy},
\end{equation}
where $H_1\equiv H$ for (\ref{1.20}), (i), (iii), and $H_1\equiv H-2\pi I$
for (\ref{1.20}), (ii). Matching Eq. (\ref{1.40}) to an analogous solution
in the adjacent cell $R_{n-1}$ leads to the solvability condition
\begin{equation}
\label{1.41}\frac{d\phi (y)}{dy}=8\pi ej_0p{
\int\limits_{L_{y1}}^{y}}duf^2(u)\sin \phi (u)+2epH_1.
\end{equation}
Equation (\ref{1.41}) is nothing but an analogue of the Ferrell-Prange\cite
{FP63} relation for a single Josephson junction, which can be readily
verified by differentiation. From this point of view, the quantity $(8\pi
ej_0p)^{-1/2}$ should be identified with the Josephson penetration depth $%
\lambda _J$. Note that instead of the factor $2\lambda $ entering the
definition of $\lambda _J$ in the single-junction case,\cite{KY72,BP82} in
our case we get the period $p$.

With the help of Eq. (\ref{1.40}), we arrive at the expression for the
vector potential in the whole region $R=
{\cup _{n=-\infty }^{+\infty }}R_n=R_s\cup R_b$:
\begin{equation}
\label{1.42}A(x,y)=\frac 1{2ep}\frac{d\phi (y)}{dy}x,\quad (x,y)\in R.
\end{equation}
This equation should be substituted into Eqs. (\ref{1.38}), (\ref{1.39}).

In this manner, we obtain a closed, complete set of equations describing a
thin-layer S/I superlattice in an external parallel magnetic field:
\begin{equation}
\label{1.43}\Delta (x,y)=\Delta _\infty f(y){\sum _n}\delta
_{R_{s_n}}(x,y)\exp \left[ in\phi (y)\right] ,
\end{equation}
$$
\delta _{R_{s_n}}(x,y)=\left\{
\begin{array}{c}
1,
\text{ for }(x,y)\in R_{s_n}, \\ 0,\text{ for }(x,y)\not \in R_{s_n};
\end{array}
\right.
$$
$$
\left[ 1-\frac 1{12}\zeta ^2(T)\left( \frac ap\right) ^2\left[ \frac{d\phi
(y)}{dy}\right] ^2\right] f(y)+\zeta ^2(T)\frac{d^2f(y)}{dy^2}-f^3(y)
$$
\begin{equation}
\label{1.44}-\frac{\alpha \zeta ^2(T)}{a\xi _0}\left[ 1-\cos \phi (y)\right]
f(y)=0,
\end{equation}
\begin{equation}
\label{1.45}\frac{df}{dy}(L_{y1})=\frac{df}{dy}(L_{y2})=0,
\end{equation}
\begin{equation}
\label{1.46}\frac{d^2\phi (y)}{dy^2}=\frac{f^2(y)}{\lambda _J^2}\sin \phi
(y),
\end{equation}
\begin{equation}
\label{1.47}\lambda _J=\left( 8\pi ej_0p\right) ^{-1/2},
\end{equation}
\begin{equation}
\label{1.48}h(y)=\frac 1{2ep}\frac{d\phi (y)}{dy},
\end{equation}
\begin{equation}
\label{1.49}j(y)\equiv j_x(y)\equiv j_0f^2(y)\sin \phi (y)=\frac 1{4\pi }
\frac{dh(y)}{dy},
\end{equation}
with boundary conditions (\ref{1.20}) and $I\equiv
{\int\limits_{L_{y1}}^{L_{y2}}}dyj(y)$, where $j(y)$ is the $x$ component
of the
supercurrent density (both in the S-layers and the barriers). The $y$
component of the intralayer supercurrent, whose average over the layer
thickness is equal to zero, within the accepted accuracy enters the theory
only implicitly, via the average kinetic-energy term in Eq. (\ref{1.44}).

Significantly, the coherent phase difference $\phi $ (the same for all the
barriers) obeys only one nonlinear second-order differential equation (\ref
{1.46}) with only one length scale, the Josephson penetration depth $\lambda
_J$, as in the case of a single junction.\cite{KY72,BP82} Due to the factor $%
f^2$, equation (\ref{1.46}) is coupled to nonlinear second-order
differential equation (\ref{1.44}), describing the spatial dependence of the
superconducting order parameter $f$ (the same for all the S-layers). In the
latter equation, the term proportional to $a^2/p^2$ accounts for the average
kinetic energy of the intralayer currents, while the term proportional to $%
\alpha $ accounts for the kinetic energy of the interlayer Josephson
currents. The Maxwell equations (\ref{1.48}), (\ref{1.49}), combined
together, yield Eq. (\ref{1.46}), as they should by virtue of
self-consistency.

It is instructive to compare the above equations with those now circulating
in literature concerned with the phenomenological LD model. As already
mentioned, neither mutual dependence of the Euler-Lagrange equations for $%
\phi _n$ and ${\bf A}$, nor fundamental complementary relations of the type (%
\ref{1.28}), (\ref{1.29}), minimizing the free energy, have been established
in previous publications. Left with an incomplete set of equations, some
authors make a non-self-consistent approximation $f_n=1$ and, regarding the
phase differences, propose a mathematically ill-defined infinite set of
differential equations with two different length scales. [See, e. g., Refs. (%
\cite{BC91,BCG92}).] In view of the conditions (\ref{1.32}), these equations
reduce to our Eq. (\ref{1.46}) with $f=1$.

Finally, the free-energy functional (\ref{1.5}) in the thin-layer limit
after a transition to the mean-field approximation with respect to ${\bf A}$
takes the form

$$
\Omega \left[ f,\phi ;H\right] =\frac{H_c^2(T)}{4\pi }W_xW_z
{\int\limits_{L_{y1}}^{L_{y2}}}dy\left[ \frac ap\left[ -f^2(y)+\frac
12f^4(y)+\zeta ^2(T)\left[ \frac{df(y)}{dy}\right] ^2\right. \right.
$$

$$
\left. +\frac{\zeta ^2(T)}{12}\left( \frac ap\right) ^2\left[ \frac{d\phi
(y) }{dy}\right] ^2f^2(y)+\frac{\alpha \zeta ^2(T)}{a\xi _0}\left[ 1-\cos
\phi (y)\right] f^2(y)\right]
$$

\begin{equation}
\label{1.50}\left. +4e^2\zeta ^2(T)\lambda ^2(T)\left[ \frac 1{2ep}\frac{%
d\phi (y)}{dy}-H\right] ^2\right] ,
\end{equation}
where $W_x=L_{x2}-L_{x1}$. As expected, minimizing Eq. (\ref{1.50}) with
respect to $f$ and the phase difference $\phi $, and neglecting terms of
order $a^2/\lambda ^2$, we arrive at Eqs. (\ref{1.44})-(\ref{1.46}).

The functional (\ref{1.50}) and complementing Maxwell equations (\ref{1.49}%
), (\ref{1.50}) contain much more physical information than the
phenomenological LD model in a parallel field: the domain of validity is
exactly determined, all the coefficients are microscopically defined, and a
finite S-layer thickness is explicitly taken into account. (As we show in
Sec. III, this factor removes unphysical divergence of $H_{c2\infty }$,
typical\cite{T96} of the LD model.) Another important difference is the
proportionality of the condensation energy in (\ref{1.50}) to the layer
thickness $a$, instead of the period $p$ in the LD functional.

The equations of the thin-layer limit admit exact solutions for all physical
situations of interest. These solutions are the subject of the next section.

\section{MAJOR PHYSICAL EFFECTS IN THE THIN-LAYER LIMIT}

\subsection{The Meissner state in a semi-infinite multilayer. The
superheating field $H_s=(ep\lambda _J)^{-1}$}

Consider a semi-infinite (in the $y$ direction) multilayer with $L_{y1}=0$, $%
L_{y2}=+\infty $ in the external fields
\begin{equation}
\label{2.1}0\leq H\leq H_s=(ep\lambda _J)^{-1},
\end{equation}
with boundary conditions of the type (\ref{1.20}), (iii):
\begin{equation}
\label{2.2}h(0)=\frac 1{2ep}\frac{d\phi }{dy}(0)=H,\text{ }h(+\infty )=\frac
1{2ep}\frac{d\phi }{dy}(+\infty )=H+4\pi I=0,\text{ }\phi (+\infty )=0.
\end{equation}
For
\begin{equation}
\label{2.3}\frac{\alpha \zeta ^2(T)}{a\xi _0}\ll 1,
\end{equation}
the Meissner solutions of Eqs. (\ref{1.44})-(\ref{1.49}) are
\begin{equation}
\label{2.4}\phi (y)=-4\arctan \frac{H\exp \left[ -\frac y{\lambda _J}\right]
}{H_s+\sqrt{H_s^2-H^2}},
\end{equation}
\begin{equation}
\label{2.5}h(y)=\frac{2HH_s\left[ H_s+\sqrt{H_s^2-H^2}\right] \exp \left[
-\frac y{\lambda _J}\right] }{\left[ H_s+\sqrt{H_s^2-H^2}\right] ^2+H^2\exp
\left[ -\frac{2y}{\lambda _J}\right] },
\end{equation}
$$
j(y)=-\frac{HH_s}{2\pi \lambda _J}\left[ H_s+\sqrt{H_s^2-H^2}\right]
$$
\begin{equation}
\label{2.6}\times \frac{\left[ \left[ H_s+\sqrt{H_s^2-H^2}\right] ^2-H^2\exp
\left[ -\frac{2y}{\lambda _J}\right] \right] \exp \left[ -\frac y{\lambda
_J}\right] }{\left[ \left[ H_s+\sqrt{H_s^2-H^2}\right] ^2+H^2\exp \left[ -
\frac{2y}{\lambda _J}\right] \right] ^2},
\end{equation}
\begin{equation}
\label{2.7}f(y)=1-\frac{4\alpha \zeta ^2(T)}{a\xi _0}\frac{H^2\left[ H_s+
\sqrt{H_s^2-H^2}\right] ^2\exp \left[ -\frac{2y}{\lambda _J}\right] }{\left[
\left[ H_s+\sqrt{H_s^2-H^2}\right] ^2+H^2\exp \left[ -\frac{2y}{\lambda _J}%
\right] \right] ^2}.
\end{equation}
The Meissner solutions persist up to the field $H_s=(ep\lambda _J)^{-1}$
that should be regarded as the superheating field of the Meissner state.

Indeed, as we will show below, the presence of Josephson vortices inside an
infinite multilayer becomes energetically favorable at a field $%
H=H_{c1\infty }<H_s$. As in the case of the well-known Bean-Livingston
barrier\cite{BL64,dG66,A88} in semi-infinite type-II superconductors, the
penetration of Josephson vortices at fields $H_{c1\infty }\leq H<H_s$ is
prevented by a surface barrier due to surface currents $j(0)$. [Compare the
discussion of the superheating field in the case of a singe junction,\cite
{KY72} where it is given by the expression $H_s=(2e\lambda \lambda _J)^{-1}$%
]. Equation (\ref{2.6}) shows that $\left| j(0)\right| $ increases in the
interval $0\leq H<H_s/\sqrt{2},$ reaches its maximum value at $H=H_s/\sqrt{2}
$, decreases in the interval $H_s/\sqrt{2}<H<H_s$ and vanishes at $H=H_s$.
Moreover, the phase difference at the surface $\phi (0)$, being a
non-positive, monotonously decreasing function of $H$ in the whole interval $%
0\leq H\leq H_s$, also reaches its minimum value $\phi (0)=-\pi $ at $H=H_s$%
. The appearance of the phase difference -$\pi $ can be attributed to the
formation of a line singularity of the amplitude of condensation $%
\left\langle \psi _{\uparrow }({\bf r})\psi _{\downarrow }({\bf r}%
)\right\rangle $ at the outer interface of the barrier (''the Josephson
vortex core''). In addition, the magnetic flux per ''elementary cell'' at $%
H=H_s$ is $\Phi =$$\Phi _0/2$, where $\Phi _0=\pi /e$ is the flux quantum.

Finally, from the second of Eqs. (\ref{2.2}) and the condition $H\leq H_s$
for the Meissner solutions, the maximal value of the total Josephson current
$\left| I\right| =I_{\max }$ in a semi-infinite multilayer is
\begin{equation}
\label{2.8}I_{\max }=\frac{H_s}{4\pi }=2\lambda _Jj_0.
\end{equation}
For $\left| I\right| >I_{\max }$, the field at the boundary is $h(0)>H_s$,
and the stationary flow of the Josephson current is disrupted by the
penetration of Josephson vortices that move under the influence of the
Lorentz force.

Thus, in fields $H>H_s$, only vortex solutions are possible. Owing to the
specific feature of the thin-layer limit, i. e., the absence of screening by
the intralayer currents, the ''tails'' of magnetic field distribution of
individual Josephson vortices overlap in the layering direction, causing the
formation of unique vortex structures that we term here ''Josephson vortex
planes''. We begin the discussion of these structures form a single ''vortex
plane'', forming in an infinite layered superconductor at the lower critical
field $H=H_{c1\infty }.$

\subsection{The lower critical field $H_{c1\infty }=2(\pi e\lambda _Jp)^{-1}$
in infinite multilayers. Vortex planes}

Consider now an infinite (in the $y$ direction) layered superconductor with $%
L_{y1}=-\infty $, $L_{y2}=+\infty $, subject to boundary conditions of the
type (\ref{1.20}), (i), with $H=0$. The condition (\ref{2.3}) is supposed to
be fulfilled. We are looking for a vortex solution with one flux quantum $%
\Phi _0$ per ''elementary cell'', i. e., with
\begin{equation}
\label{2.9}\phi (+\infty )-\phi (-\infty )=2\pi ,\text{ }\frac 1{2ep}\frac{%
d\phi }{dy}(\pm \infty )=0,\text{ }\phi (0)=\pi .
\end{equation}

The sought solution has the form of a kink:
\begin{equation}
\label{2.10}\phi (y)=4\arctan \exp \left[ \frac y{\lambda _J}\right] .
\end{equation}
This solution describes a single vortex plane positioned at $y=0$. [Compare
the phase difference $\phi (0)=\pi $ of Eq. (\ref{2.10}) with the phase
difference $\phi (0)=-\pi $ of Eq. (\ref{2.4}) at the surface of a
semi-infinite superconductor in the field $H=H_s$, when a vortex plane only
starts to penetrate. After the actual penetration, the phase difference
changes by $2\pi $, as expected from general considerations.\cite{An66}]

Corresponding distribution of the local magnetic field is given by
\begin{equation}
\label{2.11}h(y)=(ep\lambda _J)^{-1}\cosh {}^{-1}\left[ \frac y{\lambda
_J}\right] .
\end{equation}
Notice that at the vortex plane $h(0)=H_s.$ The density of the Josephson
currents is
\begin{equation}
\label{2.12}j(y)=-2j_0\cosh {}^{-2}\left[ \frac y{\lambda _J}\right] \sinh
{}\left[ \frac y{\lambda _J}\right] .
\end{equation}
At the vortex plane, $j(0)=0.$ The Josephson currents vanish exponentially
at $y\rightarrow \pm \infty $ and reach their peak values at $y=\pm \ln (1+
\sqrt{2})\lambda _J\approx \pm 0.88\lambda _J$. As regards the order
parameter, we get
\begin{equation}
\label{2.13}f(y)=1-\frac{4\alpha \zeta ^2(T)}{a\xi _0}\frac{\exp \left[ -
\frac{2y}{\lambda _J}\right] }{\left[ 1+\exp \left[ -\frac{2y}{\lambda _J}%
\right] \right] ^2}.
\end{equation}

Notice that Eqs. (\ref{2.11})-(\ref{2.13}), considered in the half-space $%
0\leq y<+\infty $, have exactly the same form as the solutions (\ref{2.5})-(%
\ref{2.7}) for a semi-infinite multilayer in the external field $H=H_s$, in
full agreement with our interpretation of $H_s$ as the penetration field for
a single vortex plane.

To find the lower critical field $H_{c1\infty }$ at which the solution (\ref
{2.10}) becomes energetically favorable, we must consider the free-energy
functional (\ref{1.50}), which in this case takes the form%
$$
\Omega \left[ \phi (y);H\right] -\Omega \left[ H\right]
_{N_v=0}=N_{cell}W_z\left[ \frac{j_0}{2e}
{\int\limits_{-\infty }^{+\infty }}dy\left[ 1-\cos \phi (y)+\frac{\lambda
_J^2}2\left[ \frac{d\phi (y)}{dy}\right] ^2\right] \right.
$$
\begin{equation}
\label{2.14}\left. -\frac 1{4\pi }\frac{\left[ \phi (+\infty )-\phi (-\infty
)\right] H}{2e}\right] ,
\end{equation}
where $\Omega \left[ H\right] _{N_v=0}\equiv \Omega \left[ \phi =0;H\right] $
is the free energy in the absence of vortices ($N_v$ is the number of vortex
planes), and $N_{cell}=W_x/p$ is the number of ''elementary cells''.
Inserting (\ref{2.10}) into (\ref{2.14}), we obtain the free-energy
contribution of a single vortex plane:
\begin{equation}
\label{2.15}\Omega \left[ H\right] _{N_v=1}-\Omega \left[ H\right]
_{N_v=0}=N_{cell}W_z\left[ \frac{4\lambda _Jj_0}e-\frac{\Phi _0H}{4\pi }%
\right] ,
\end{equation}
with $\Phi _0=\pi /e$ the flux quantum. From the condition $\Omega \left[
H_{c1\infty }\right] _{N_v=1}=\Omega \left[ H_{c1\infty }\right] _{N_v=0}$,
the lower critical field is
\begin{equation}
\label{2.16}H_{c1\infty }=2(\pi ep\lambda _J)^{-1}=\frac 2\pi \frac{\Phi _0}{%
\pi p\lambda _J},
\end{equation}
as in the case of a single junction, apart from the factor $p$ in the
denominator instead of $2\lambda (T)$.\cite{KY72,BP82} As expected, $%
H_{c1\infty }=2H_s/\pi <H_s=h(0)$. On the contrary, Eq. (\ref{2.16}) is
completely different from previously proposed ones for layered
superconductors,\cite{B73} based on an invalid assumption of single-vortex
penetration.

From the proportionality of the right-hand side of (\ref{2.15}) to $N_{cell}$%
, we infer that the total number of Josephson vortices (i. e., 1D
singularities of the amplitude of condensation) in one vortex plane is equal
to the total number of elementary cells. This means that Josephson vortices
penetrate all the cells simultaneously and coherently. As in the case of a
single junction,\cite{KY72} the quantity
\begin{equation}
\label{2.17}E_0=\frac{4\lambda _Jj_0}e
\end{equation}
can be identified with the self-energy of a single Josephson vortex per unit
length (in the $z$ direction).

In higher external fields ($H\gg H_{c1\infty }$), we expect to get a
''stack'' of $N_v$ vortex planes with the total number of Josephson vortices
$N_{v_{tot}}=N_vN_{cell}$. (See Fig. 2.)

\subsection{The vortex state in intermediate fields $H_{c1\infty }\ll H\ll
\left[ ea\zeta (T)\right] ^{-1}$. The lower critical field $H_{c1W}=\pi /epW$
in finite-size samples ($W\ll \lambda _J$). A paramagnetic effect}

Consider a finite-size (in the $y$ direction) multilayer with $%
-L_{y1}=L_{y2}\equiv L,$ $W\equiv 2L$, in the field range $H_{c1\infty }\ll
H\ll \left[ ea\zeta (T)\right] ^{-1}$ and in the absence of externally
applied current ($I=0$), i. e., subject to the boundary conditions (\ref
{1.20}), (i). The validity of the condition (\ref{2.3}) is again assumed.
[The upper bound $\left[ ea\zeta (T)\right] ^{-1}$ for the field range means
that we are concerned with $H\ll H_{c2\infty }(T)$.]

Under these assumptions, the phase difference up to first order in the small
parameter $(ep\lambda _JH)^{-2}$ is

\begin{equation}
\label{2.18}\phi (y)=2epHy+\pi N_v(H)-\frac{(-1)^{N_v}}{4(ep\lambda _JH)^2}%
\left[ \sin \left( 2epHy\right) -2epHy\cos \left( epWH\right) \right] .
\end{equation}
The constant of integration $\pi N_v(H)$ accounts here for the phase shift
due to $N_v$ vortex planes [$\pi $ per each vortex plane, see the last of
Eqs. (\ref{2.9})]. The number of vortex planes $N_v$ is itself a singular
function of the applied field $H$:
\begin{equation}
\label{2.19}N_v(H)=\left[ \frac{epWH}\pi \right] =\left[ \frac \Phi {\Phi
_0}\right] ,
\end{equation}
where $\left[ u\right] $ means the integer part of $u$, and $\Phi =pWH$ is
the flux through an ''elementary cell''. This choice of the constant of
integration guarantees that the energy of the Josephson coupling $E_J$ in (%
\ref{1.50}) takes the minimal value for a given $H$:
\begin{equation}
\label{2.20}E_J\left[ H\right] =\frac{H_c^2(T)}{4\pi }W_xWW_z\frac{\alpha
\zeta ^2(T)}{a\xi _0}\left[ 1-\frac \Phi {\Phi _0}\left| \sin \frac{\pi \Phi
}{\Phi _0}\right| \right] .
\end{equation}
(This expression should be compared with its analog for a single Josephson
junction.\cite{KY72})

The physical quantities corresponding to (\ref{2.18}) are
\begin{equation}
\label{2.21}h(y)=H\left[ 1-\frac{(-1)^{N_v}}{4(ep\lambda _JH)^2}\left[ \cos
\left( 2epHy\right) -\cos \left( epWH\right) \right] \right] ,
\end{equation}
\begin{equation}
\label{2.22}j(y)=(-1)^{N_v}j_0\sin \left( 2epHy\right) ,
\end{equation}
\begin{equation}
\label{2.23}f(y)=1-\frac{\alpha \zeta ^2(T)}{2a\xi _0}\left[ 1-\frac{%
(-1)^{N_v}\cos \left( 2epHy\right) }{1+2\left[ epH\zeta (T)\right] ^2}-\frac{
\sqrt{2}epH\zeta (T)\left| \sin \left( epHW\right) \right| }{1+2\left[
epH\zeta (T)\right] ^2}\frac{\cosh \frac{\sqrt{2}y}{\zeta (T)}}{\sinh \frac
W{\sqrt{2}\zeta (T)}}\right] .
\end{equation}
[The term $2\left[ epH\zeta (T)\right] ^2$ in the denominators of (\ref{2.23}%
) can only be retained if $p\gg a$.]

In the limit $W\gg \zeta (T)$, $\left| y\right| \ll W/2$, equation (\ref
{2.23}) becomes
\begin{equation}
\label{2.24}f(y)=1-\frac{\alpha \zeta ^2(T)}{2a\xi _0}\left[ 1-\frac{%
(-1)^{N_v}\cos \left( 2epHy\right) }{1+2\left[ epH\zeta (T)\right] ^2}%
\right] .
\end{equation}
Equations of the type (\ref{2.18}), (\ref{2.21}) and (\ref{2.24}) for $%
N_v=2m $ ($m$ is an integer) were first obtained by Theodorakis\cite{Th90}
in the framework of the LD model. Our equation (\ref{2.18}) for $N_v=2m$
should also be compare with an analogous solution for an infinite single
junction given, for instance, in Ref. (\cite{A88}).

The singular function $N_v(H)$ introduces discontinuities in Eqs. (\ref{2.18}%
), (\ref{2.21})-(\ref{2.24}). These discontinuities witness that the system
undergoes a first-order phase transition when a vortex plane penetrates or
leaves the sample. [Compare with the discussion of a single junction in Ref.
(\cite{KY72}).]

The positions of vortex planes $y_v$ correspond to local maxima of the field
$h(y)$ in Eq. (\ref{2.21}). [In the case (\ref{2.24}), $y_v$ exactly
coincide with local minima of $f(y)$.] In the vortex planes $y=y_v$, the
microscopic magnetic field is higher than the applied one:
\begin{equation}
\label{2.25}h(y_v)=H\left[ 1+\frac 1{4(ep\lambda _JH)^2}\left[
1+(-1)^{N_v}\cos \left( epWH\right) \right] \right] >H,
\end{equation}
which is expected for any vortex solution. The Josephson current density $%
j(y)=\frac 1{4\pi }\frac{dh(y)}{dy}$ vanishes both in the vortex planes $%
y=y_v$ and in the planes of local minima of $h(y)$, $y=y_v\pm \pi /2epH$.
When passing through zero in these planes, $j(y)$ changes the sign, as
depicted in Fig. 2.

From Eq. (\ref{2.19}) with $N_v(H)=1$, we obtain the lower critical field $%
H_{c1W}$ in a finite multilayer with $W\ll \lambda _J$:
\begin{equation}
\label{2.26}H_{c1W}=\frac \pi {epW}=\frac{\pi ^2}2H_{c1\infty }\frac{\lambda
_J}W\gg H_{c1\infty }.
\end{equation}

The definition of the magnetization $\dot M$,\cite{A88}
\begin{equation}
\label{2.27}4\pi M=\frac 1W{\int\limits_{-L}^{+L} }dyh(y)-H
\end{equation}
and Eq. (\ref{2.21}) yield
\begin{equation}
\label{2.28}M(H)=-\frac 1{16\pi H(ep\lambda _J)^2}\left[ \frac{\left| \sin
\left( epWH\right) \right| }{epWH}-(-1)^{N_v}\cos \left( epWH\right) \right]
.
\end{equation}
The magnetization (\ref{2.28}) shows distinctive oscillatory behavior and
discontinuities at $epWH\rightarrow \pi N$ ($N$ is an integer), when a
vortex plane penetrates or leaves the sample.

Interestingly enough, the right-hand side of (\ref{2.28}) passes through
zero and may have both signs. Thus, for $\Phi =pWH\gg \Phi _0$, the sample
exhibits a small paramagnetic effect, if $N_v\Phi _0<$$\Phi <\left(
N_v+\frac 12-\frac{\Phi _0}{\pi ^2\Phi }\right) \Phi _0:$%
\begin{equation}
\label{2.29}M(H)=\frac 1{16\pi H(ep\lambda _J)^2}\left[ \left| \cos \left(
epWH\right) \right| -\frac{\left| \sin \left( epWH\right) \right| }{epWH}%
\right] >0.
\end{equation}

\subsection{Fraunhofer oscillations of the Josephson current in multilayers
with $W\ll \lambda _J$, in the field range $0\leq H\ll \left[ ea\zeta
(T)\right] ^{-1}$. ''Edge pinning'' of the vortex planes}

Now we proceed to the case of a finite-size (along the layers) multilayer
with $-L_{y1}=L_{y2}\equiv L,$ $W\equiv 2L$ in the presence of an externally
applied current $I$, i. e., subject to the boundary conditions (\ref{1.20}),
(ii). (Compare the discussion by Owen and Scalapino\cite{OS67} of the
single-junction case.) The relation (\ref{2.3}) is supposed to hold. The
applied magnetic fields are within the range $0\leq H\ll \left[ ea\zeta
(T)\right] ^{-1}$.

Assuming $W\ll \lambda _J$, we can consider $W^2/\lambda _J^2$ as a small
expansion parameter in Eq. (\ref{1.46}). In this way, we obtain%
$$
\phi (y)=2epHy+\pi N_v(H)+\varphi
$$
\begin{equation}
\label{2.30}-\frac{(-1)^{N_v}}4\frac{W^2}{\lambda _J^2}\left[ \frac{\Phi _0}{%
\pi \Phi }\right] ^2\left[ \sin \left( 2epHy+\varphi \right) -2epHy\cos
\frac{\pi \Phi }{\Phi _0}\cos \varphi -\sin \varphi \right] ,
\end{equation}
\begin{equation}
\label{2.31}I(\varphi ,\Phi )\equiv {\int\limits_{-L}^{+L} }%
dyj(y)=j_0W\frac{\Phi _0}{\pi \Phi }\left| \sin \frac{\pi \Phi }{\Phi _0}%
\right| \sin \varphi ,
\end{equation}
\begin{equation}
\label{2.32}h(y)=H\left[ 1-\frac{(-1)^{N_v}}4\frac{W^2}{\lambda _J^2}\left[
\frac{\Phi _0}{\pi \Phi }\right] ^2\left[ \cos \left( 2epHy+\varphi \right)
-\cos \frac{\pi \Phi }{\Phi _0}\cos \varphi \right] \right] ,
\end{equation}
$$
f(y)=1-\frac{\alpha \zeta ^2(T)}{a\xi _0}\left[ 1-\frac{(-1)^{N_v}\cos
\left( 2epHy+\varphi \right) }{1+2\left[ epH\zeta (T)\right] ^2}\right.
$$
$$
-\frac{\sqrt{2}epH\zeta (T)\left| \sin \left( epHW\right) \right| \cos
\varphi }{1+2\left[ epH\zeta (T)\right] ^2}\frac{\cosh \frac{\sqrt{2}y}{%
\zeta (T)}}{\sinh \frac W{\sqrt{2}\zeta (T)}}
$$
\begin{equation}
\label{2.33}\left. -\frac{(-1)^{N_v}\sqrt{2}epH\zeta (T)\cos \left(
epHW\right) \sin \varphi }{1+2\left[ epH\zeta (T)\right] ^2}\frac{\sinh
\frac{\sqrt{2}y}{\zeta (T)}}{\cosh \frac W{\sqrt{2}\zeta (T)}}\right] ,
\end{equation}
where $N_v(H)=\left[ \frac{epWH}\pi \right] $ is the number of vortex
planes, $\Phi =pWH$ is the flux through an ''elementary cell'', $\Phi _0=\pi
/e$, as usual, and the constant $\varphi $ ($\left| \varphi \right| \leq \pi
/2$) parameterizes the total Josephson current $I$ given by (\ref{2.31}).
Equation (\ref{2.31}) yields the well-known Fraunhofer pattern, the only
difference from the single-junction case being the occurrence of the period $%
p$ in place of $2\lambda (T)$.\cite{KY72,BP82} Note that the first zero of
the Fraunhofer pattern, by virtue of (\ref{2.26}), corresponds to the lower
critical field $H_{c1W}$. In the absence of the transport current, i. e.,
for $\varphi =0$, equations (\ref{2.30}), (\ref{2.32}), (\ref{2.33}) reduce,
respectively, to (\ref{2.18}), (\ref{2.21}) and (\ref{2.23}), as they should.

The self-consistency of our calculations can be easily verified by means of
Ampere's law $h(+L)-h(-L)=4\pi I$. It is assured by terms proportional to $%
W^2/\lambda _J^2$ in (\ref{2.30}) and (\ref{2.32}) that explicitly take into
account the effect of self-induced fields. Although Eq. (\ref{2.31}) was
first derived in the framework of the LD model in Ref. (\cite{BCG92}), the
authors of that publication, based on an incomplete set of equations, were
unable to find the phase differences self-consistently and evaluate the
local magnetic field in first order in $W^2/\lambda _J^2$. As a result, they
arrived at an absolutely incorrect conclusion that Fraunhofer oscillations
of $I$ could be observed in the absence of Josephson vortices.
Unfortunately, this misunderstanding is shared in some other recent
publications.\cite{FG94} Therefore, we provide below a detailed and rigorous
clarification.

As we see from Eq. (\ref{2.32}), in the presence of the transport current $I$%
, the vortex planes are shifted by the Lorentz force to new equilibrium
positions [local maxima of $h(y)]:$%
\begin{equation}
\label{2.34}\bar y_v=y_v-\frac \varphi {2epH},
\end{equation}
where $y_v$ correspond to local maxima of the right-hand side of (\ref{2.32}%
) for $\varphi =0$. The local magnetic field in the vortex planes now is
\begin{equation}
\label{2.35}h(\bar y_v)=H\left[ 1+\frac 1{4(ep\lambda _JH)^2}\left[
1+(-1)^{N_v}\cos \left( epWH\right) \cos \varphi \right] \right] >H.
\end{equation}

In equilibrium, the Lorentz force $f_L$ per elementary cell acting on the
vortex planes is counterbalanced by the pinning force $f_{pin}$ that can be
defined as\cite{CE72}
\begin{equation}
\label{2.36}f_{pin}(Y)=-\frac 1{N_{cell}W_z}\frac{dU_{pin}(Y)}{dY},
\end{equation}
where $U_{pin}(Y)$ is the pinning potential arising owing to the shift by $Y$
of the vortex planes from their equilibrium positions in the absence of the
transport current $I$. To evaluate the pinning potential, we consider the
increase of the free energy in first order in $\alpha $$\zeta ^2(T)/a\xi _0,$
caused by such a shift. Noting that first-order corrections to $f(y)\approx
1 $ and $h(y)\approx H$ do not contribute to the free energy, taking $\phi
(y)\approx 2epHy+\pi N_v(H)$, making the transformation $y\rightarrow y-Y$
and substituting into Eq. (\ref{1.50}), we obtain%
$$
U_{pin}(Y)\equiv \Omega \left[ H;Y\right] -\Omega \left[ H;Y=0\right]
$$
\begin{equation}
\label{2.37}=N_{cell}WW_z\frac{j_0}{2e}\frac{\Phi _0}{\pi \Phi }\left| \sin
\frac{\pi \Phi }{\Phi _0}\right| \left[ 1-\cos \left( \frac{2\pi \Phi }{\Phi
_0}\frac YW\right) \right] .
\end{equation}
It is very instructive to rewrite Eq. (\ref{2.37}) as
\begin{equation}
\label{2.38}U_{pin}(Y)=N_{cell}WW_z\frac 1{2e}\frac{\Phi _0}{\pi \Phi }%
\left[ 2j(+L;Y=0)+j(-L;Y)-j(+L;Y)\right] ,
\end{equation}
where%
$$
j(\pm L;Y)=(-1)^{N_v}j_0\sin (\pm 2epHL+2epHY)
$$
are the surface currents in the presence of the shift $Y$. We see that the
pinning potential for $\left| Y\right| <\pi /4epH$ arises owing to the
emergence of additional surface currents on the opposite sides of the
superconductor. At $2epHL=\pi N+\frac 12$ ($N$ is an integer), $%
j(-L;Y)=-j(+L;Y)$, i. e. these currents flow in the opposite directions, and
the pinning potential reaches its maximum. On the contrary, at $2epHL=\pi N$
($N$ is an integer), $j(-L;Y)=j(+L;Y)$, i. e. the surface currents flow in
the same direction and mutually compensate each other in (\ref{2.38}), the
pinning potential vanishes, and vortex planes freely penetrate or leave the
sample. (Compare with the discussion at the beginning of this Section of the
case of a semi-infinite multilayer for $H=H_s$.) The surface currents also
flow in the same direction and mutually compensate each other when the
magnitude of the shift $\left| Y\right| $ reaches the value $\left| Y\right|
=Y_{\max }\equiv \pi /4epH$. Moreover, the pinning potential vanishes for $%
\Phi \gg \Phi _0$.

From (\ref{2.36}) and (\ref{2.37}), we obtain the pinning force for the
shift $Y$:
\begin{equation}
\label{2.39}f_{pin}(Y)=-I\left( \frac{2\pi \Phi }{\Phi _0}\frac YW;\Phi
\right) \Phi ,
\end{equation}
$$
I\left( \frac{2\pi \Phi }{\Phi _0}\frac YW;\Phi \right) \equiv j_0W\frac{%
\Phi _0}{\pi \Phi }\left| \sin \frac{\pi \Phi }{\Phi _0}\right| \sin \left(
\frac{2\pi \Phi }{\Phi _0}\frac YW\right) .
$$
From these expressions we infer that the maximal value of the pinning force $%
\left| f_{pin}\right| $ for given flux $\Phi $ is $f_{pin}^{\max }=\left|
I\left( \left| \frac \pi 2\right| ;\Phi \right) \right| \Phi $.

In the presence of the transport current $I(\varphi ;\Phi )$ [Eq. (\ref{2.31}%
)], the shift of the positions of the vortex planes, according to (\ref{2.34}%
), is $Y=-\varphi /2epH$, with the maximal equilibrium value $\left|
Y\right| =Y_{\max }\equiv \pi /4epH$. Taking into account the fact that in
equilibrium $f_L=-f_{pin}$, we arrive at the expression for the
corresponding Lorentz force:
\begin{equation}
\label{2.40}f_L=-I(\varphi ;\Phi )\Phi .
\end{equation}
This expression was to be expected from general considerations,\cite{CE72}
which prescribe for the magnitude of the Lorentz force the relation$\left|
f_L\right| =\left| I\right| \Phi $, where $I$ is the transport current. It
is therefore absolutely clear that the stationary Josephson effect becomes
impossible if the magnitude of the transport current $\left| I\right| $
exceeds the value $I_{\max }=\left| I\left( \left| \frac \pi 2\right| ;\Phi
\right) \right| $, because in this situation $\left| f_L\right| >\left|
f_{pin}\right| $, and the vortex planes are completely depinned.

Notice that the physics of the Fraunhofer pattern in single junctions was
discussed in terms of a series of first-order phase transitions due to
successive penetration of Josephson vortices long ago.\cite{KY72} A
qualitative explanation by means of the ''edge pinning'' was proposed in the
book by Tinkham.\cite{T96} In general, the pinning of Josephson vortices in
weakly coupled superconducting structures with $W\ll \lambda _J$ is
completely analogous to the pinning of Abrikosov vortices by the edges of a
thin [compared to $\lambda (T)]$ type-II superconducting film.\cite{Sh69}

Finally, we observe that the magnetization in the presence of the transport
current $I(\varphi ;\Phi )$, according to (\ref{2.27}) and (\ref{2.32}), is
given by
\begin{equation}
\label{2.41}M(H)=-\frac 1{16\pi H(ep\lambda _J)^2}\left[ \frac{\left| \sin
\left( epWH\right) \right| }{epWH}-(-1)^{N_v}\cos \left( epWH\right) \right]
\cos \varphi .
\end{equation}
For $\varphi =0$, equation (\ref{2.41}) reduces to (\ref{2.28}). For $\Phi
=pWH\gg \Phi _0$, $N_v\Phi _0<$$\Phi <\left( N_v+\frac 12-\frac{\Phi _0}{\pi
^2\Phi }\right) \Phi _0$, we again obtain the paramagnetic effect. [Compare
with Eq. (\ref{2.29}).]

\subsection{Critical parameters of an infinite multilayer: $T_{c\infty }$, $%
H_{c2\infty }$}

At the point of the second-order phase transition to the normal state, $f^2$
can be considered as a small parameter. Thus, the term $f^3$ in Eq. (\ref
{2.19}) and the right-hand side of Eq. (\ref{2.21}) can be dropped. Applying
boundary conditions (\ref{1.20}), (i) yields $\phi (y)=2epHy+\pi N_v(H)$.
With this phase difference, the linearized version of (\ref{1.19}) can be
transformed into
\begin{equation}
\label{2.42}\frac{d^2f(t)}{dt^2}+\left[ A(T,H)-(-1)^{N_v+1}q(H)\cos
2t\right] f(t)=0,
\end{equation}
$$
A(T,H)\equiv \frac{1-\frac 13e^2H^2a^2\zeta ^2(T)-\frac{\alpha \zeta ^2(T)}{%
a\xi _0}}{\left[ ep\zeta (T)H\right] ^2},\text{ }q(H)\equiv \frac \alpha
{2a\xi _0\left( epH\right) ^2},
$$
where we have introduced a dimensionless variable $t\equiv epHy$: $%
f(t)\equiv f(t/epH)$. Hence one gets two independent equations: for the odd $%
N_v=2m+1$ ($m=0,1,2,\ldots $) and the even $N_v=2m$ number of vortex planes.
Both of them have the usual form of Mathieu equations. [See Appendix A.] As
to the boundary conditions, it is convenient to take $-L_{y1}=L_{y2}\equiv
L\equiv W/2$ and, by symmetry, consider Eq. (\ref{2.42}) in the interval $%
0\leq y\leq L$, with
\begin{equation}
\label{2.43}\frac{df}{dt}(0)=\frac{df}{dt}(epHL)=0.
\end{equation}
The critical parameters $T_c$ and $H_{c2}$ are now determined by the
smallest eigenvalue of the boundary problem (\ref{2.42}), (\ref{2.43}).

In an infinite in the $y$ direction multilayer ($L\rightarrow \infty $), the
only bounded at the infinity solutions of (\ref{2.42}) are periodic Mathieu
functions, with $f_{N_v=2m+1}(t)\propto $ce$_0(t,q)$ and $%
f_{N_v=2m}(t)\propto $ce$_0(\pi /2-t,q)$ corresponding to the smallest
eigenvalues $a_0(q)$ and $a_0(-q)=a_0(q)$, respectively. Thus the critical
parameters are given by the equation
\begin{equation}
\label{2.44}\left[ A(T,H)\right] _{c\infty }=\left[ a_0(q)\right] _{c\infty
},
\end{equation}
where one should fix $H$ to obtain $T_{c\infty }$ or, alternatively, fix $T$
to obtain $H_{c2\infty }$. As in the case of Eq. (\ref{2.24}), local minima
of the reduced order parameter $f(t)$ in (\ref{2.42}) correspond to the
positions of the vortex planes: in conventional units the distance between
two successive minima is $\Delta y_v=\pi /epH$, which gives the flux $\Phi
=\Delta y_vpH=\Phi _0$ per single vortex.

\subsubsection{The critical temperature $T_{c\infty }$}

For magnetic fields $H$ in the range
\begin{equation}
\label{2.45}H\ll \frac{\Phi _0}{a\xi _0\chi ^{1/2}\left( \xi _0/l\right) },
\end{equation}
the general expression for $T_{c\infty }$ resulting from Eq. (\ref{2.44}) is
\begin{equation}
\label{2.46}T_{c\infty }=T_{c0}\left[ 1-\frac{7\zeta (3)}{12}\xi _0^2\chi
\left( \xi _0/l\right) \left[ \frac 13e^2H^2a^2+\frac \alpha {a\xi
_0}+\left( epH\right) ^2a_0\left( \frac \alpha {2a\xi _0\left( epH\right)
^2}\right) \right] \right] .
\end{equation}

In weak fields
\begin{equation}
\label{2.47}H\ll \frac 1{ep}\sqrt{\frac \alpha {2a\xi _0}}\equiv \frac{\Phi
_0}{\pi p}\sqrt{\frac \alpha {2a\xi _0}},
\end{equation}
pair-breaking effect of intralayer supercurrents is unimportant, and we get
\begin{equation}
\label{2.48}T_{c\infty }=T_{c0}\left[ 1-\frac{7\sqrt{2}\zeta (3)}{12}\xi
_0^2\chi \left( \xi _0/l\right) \sqrt{\frac \alpha {a\xi _0}}epH\right] .
\end{equation}

In strong fields
\begin{equation}
\label{2.49}\frac{\Phi _0}{\pi p}\sqrt{\frac \alpha {2a\xi _0}}\ll H\ll
\frac{\Phi _0}{a\xi _0\chi ^{1/2}\left( \xi _0/l\right) },
\end{equation}
Eq. (\ref{2.46}) becomes
\begin{equation}
\label{2.50}T_{c\infty }=T_{c0}\left[ 1-\frac{7\zeta (3)}{12}\xi _0^2\chi
\left( \xi _0/l\right) \left[ \frac 13e^2H^2a^2+\frac \alpha {a\xi
_0}\right] \right] .
\end{equation}
The term proportional to $\alpha $ takes into account pair-breaking by the
Josephson currents, locally equal to the critical ones.\cite{KN97} In the
absence of weak coupling ($\alpha =0$), equation (\ref{2.50}) reduces to the
well-known expression for an isolated thin superconducting film.\cite
{dG66,T96}

\subsubsection{The upper critical field $H_{c2\infty }$}

For a fixed $T,$ equation (\ref{2.44}) yields an implicit expression for $%
H_{c2\infty }$ as a function of $T$:
\begin{equation}
\label{2.51}\left[ ep\zeta (T)H_{c2\infty }\right] ^2\left[ \frac 13\left(
\frac ap\right) ^2+a_0\left( \frac{\alpha \zeta ^2(T)}{2a\xi _0\left[
ep\zeta (T)H_{c2\infty }\right] ^2}\right) \right] =1-\frac{\alpha \zeta
^2(T)}{a\xi _0}.
\end{equation}
This expression exhibits the so-called\cite{T96} ''3D-2D crossover'',
experimentally verified, for example, on Nb/Ge multilayers.\cite{RBB80} The
''crossover temperature'' $T^{*}$ can be conventionally defined by the
relation $\alpha $$\zeta ^2(T^{*})/a\xi _0=1$.

For temperatures close to $T_{c0}$, when
\begin{equation}
\label{2.52}\frac{\alpha \zeta ^2(T)}{a\xi _0}\gg 1,
\end{equation}
\begin{equation}
\label{2.53}H_{c2\infty }(T)=\frac{\Phi _0}{\sqrt{2}\pi p\zeta (T)}\frac{
\sqrt{a\xi _0}}{\sqrt{\alpha }\zeta (T)}=\frac{12}{7\sqrt{2}\zeta (3)\pi
\sqrt{\alpha }}\frac{\Phi _0\sqrt{a\xi _0}}{p\xi _0^2\chi \left( \xi
_0/l\right) }\left( 1-\frac T{T_{c0}}\right) .
\end{equation}
In this ''3D'' regime, the positive kinetic energy of small interlayer
Josephson currents in (\ref{1.50}) competes with the negative intralayer
condensation energy. The superconductivity of the S-layers is strongly
depressed by the vortex planes, as a comparison between local maxima $%
f_{\max }$ and local minima $f_{\min }$ of the order parameter shows:
\begin{equation}
\label{2.54}\frac{f_{\min }}{f_{\max }}\equiv \frac{f(y_v)}{f(y_v\pm \pi
/2epH_{c2\infty })}=2\sqrt{2}\exp \left[ -\frac{2\alpha \zeta ^2(T)}{a\xi _0}%
\right] \ll 1.
\end{equation}

At lower temperatures, when
\begin{equation}
\label{2.55}\frac{\alpha \zeta ^2(T)}{a\xi _0}\ll 1,
\end{equation}
\begin{equation}
\label{2.56}H_{c2\infty }(T)=\frac{\sqrt{3}\Phi _0}{\pi a\zeta (T)}\left[
1-\frac 12\frac{\alpha \zeta ^2(T)}{a\xi _0}\right] .
\end{equation}
In this ''2D'' regime, the energy of the Josephson interlayer coupling is
small relative to the intralayer condensation energy.\cite{KN97} The
transition to the normal phase occurs owing mainly to pair-breaking by the
intralayer supercurrents, and the order parameter is almost unperturbed by
the vortex planes:
\begin{equation}
\label{2.57}f(y)\propto 1-\frac{(-1)^{N_v}}{12}\frac{a^2}{p^2}\frac{\alpha
\zeta ^2(T)}{a\xi _0}\cos \left( 2epHy\right) .
\end{equation}
This expression should be compared with Eq. (\ref{2.24}) for intermediate
fields in the same temperature range (\ref{2.55}). In the limit $\alpha =0$
(no Josephson coupling), equation (\ref{2.56}) goes over into the well
familiar one for an isolated thin superconducting film.\cite{dG66,A88,T96}
Equation (\ref{2.56}) explains the origin of the well-known\cite{T96}
unphysical ''low-temperature'' divergence of the LD model: taking a formal
limit $a\rightarrow 0$ while keeping $\alpha \zeta ^2(T)/a\xi _0=$const, we
get $H_{c2\infty }(T)\rightarrow \infty $.

Aside from microscopically determined parameters, dependence (\ref{2.53})
for layered superconductors was first obtained within the framework of the
LD model.\cite{LD,KLB75,T96} Expressions of the type (\ref{2.56}) were
derived phenomenologically in Refs. (\cite{DE78,SS93}). In all these
publications relations (\ref{1.31}), (\ref{1.32}) were implicitly adopted as
physically plausible assumptions. The very fact that these results are
contained as limiting cases in our Eqs. (\ref{1.43})-(\ref{1.49}) once again
demonstrates the generality and self-consistency of the approach of this
paper.

Finally, we emphasize that the concept of Josephson vortex planes applies
both in limits (\ref{2.52}) and (\ref{2.55}). Contrary to previous
suggestions,\cite{KLB75,BCG92} there is no transition from the
''Abrikosov-core regime'' to the ''Josephson-core regime'' at $T^{*}$: The
existence of Abrikosov vortices with normal cores in the thin-layer limit is
not allowed by the solutions of Eq. (\ref{2.42}). [Mathematically, the
function ce$_0(t,q)$ is strictly positive.]

\subsection{Size-effects: Oscillations of $T_{cW}$}

Aside from a special case $epHL=\pi k/2$ ($\Phi =k\Phi _0,$ $k=0,1,2,\ldots $%
), for multilayers with finite S-layer length $W=2L$ only approximate
solutions of the boundary problem (\ref{2.42}), (\ref{2.43}) can be
obtained. Using Galerkin's method,\cite{R80} we have found two groups of
solutions corresponding to the smallest eigenvalues $\left[ A\right] _c$:
\begin{equation}
\label{2.58}f_{\mu ,N_v=2n+1}(t)\propto \cosh \left( \mu t\right) \text{ce}%
_0(t,q),\text{ }f_{\mu ,N_v=2n}(t)\propto \cosh \left( \mu t\right) \text{ce}%
_0\left( \frac \pi 2-t,q\right) ,
\end{equation}
\begin{equation}
\label{2.59}\mu =\coth \left( \mu epHL\right) \frac{\left| \frac{d\text{ce}%
_0 }{dt}\left( epHL,q\right) \right| }{\text{ce}_0\left( epHL,q\right) },
\end{equation}
\begin{equation}
\label{2.60}\left[ A(T,H)\right] _c=\left[ a_0(q)-\mu ^2\right] _c,
\end{equation}
and
\begin{equation}
\label{2.61}f_{\nu ,N_v=2n+1}(t)\propto \cos \left( \nu t\right) \text{ce}%
_0(t,q),\text{ }f_{\nu ,N_v=2n}(t)\propto \cos \left( \nu t\right) \text{ce}%
_0\left( \frac \pi 2-t,q\right) ,
\end{equation}
\begin{equation}
\label{2.62}\nu =-\cot \left( \nu epHL\right) \frac{\left| \frac{d\text{ce}%
_0 }{dt}\left( epHL,q\right) \right| }{\text{ce}_0\left( epHL,q\right) },
\end{equation}
\begin{equation}
\label{2.63}\left[ A(T,H)\right] _c=\left[ a_0(q)+\nu ^2\right] _c,
\end{equation}
where Eqs. (\ref{2.59}), (\ref{2.62}) implicitly define parameters $\mu $
and $\nu $, and Eqs. (\ref{2.60}), (\ref{2.63}) determine the critical point.

From Eqs. (\ref{2.60}), (\ref{2.63}) we see, that the eigenvalue
corresponding to the eigenfunctions $f_\mu $ is smaller than $a_0(q)$, while
that corresponding to the eigenfunctions $f_\nu $ is larger. Physically,
this means that with $f_\mu $ in a finite multilayer we can achieve higher
values of the critical parameters $T_c$ and $H_{c2}$ than in an infinite
one. [Compare with Eq. (\ref{2.44}).] At $epHL=\pi k/2$ ($\Phi =k\Phi _0,$ $%
k=0,1,2,\ldots $), these equations yield $\mu =\nu =0$.

For $epHL\rightarrow \infty $, $\mu $ is a bounded, oscillating function of $%
epHL$ and does not tend to any limit. On the contrary, $\nu \rightarrow \pi
/2epHL,$ when $epHL\rightarrow \infty $. This signifies that at certain
values of $epHL$ the solution $f_\mu $ becomes unstable and gives way to the
solution $f_\nu $ with lower values of the critical parameters, presumably
by means of a first-order phase transition. As the parameters $\mu $ and $%
\nu $ can enter the free energy only via the combinations $\mu epHL$ and $%
\nu epHL$, we expect the transitions $f_\mu \leftrightarrow f_\nu $ to occur
when $\left[ \mu epHL\right] _{*}=\left[ \nu epHL\right] _{*},$ hence the
relation
$$
\cot \left[ \mu epHL\right] _{*}=-\coth \left[ \mu epHL\right] _{*}
$$
with the numerical solution $\left[ \mu epHL\right] _{*}\approx 2.37.$ Thus,
the solution $f_\mu $ with (\ref{2.60}) is realized when
\begin{equation}
\label{2.64}\frac{epHL\left| \frac{d\text{ce}_0}{dt}\left( epHL,q\right)
\right| }{\text{ce}_0\left( epHL,q\right) }<\left[ \mu epHL\right] _{*}\tanh
\left[ \mu epHL\right] _{*}\approx 2.33,
\end{equation}
while for
\begin{equation}
\label{2.65}\frac{epHL\left| \frac{d\text{ce}_0}{dt}\left( epHL,q\right)
\right| }{\text{ce}_0\left( epHL,q\right) }>2.33,
\end{equation}
the system ''chooses'' $f_\nu $ with (\ref{2.63}). The condition (\ref{2.64}%
) is met, for instance, when $\Phi \equiv pWH\approx k\Phi _0$ ($W=2L,$ $%
k=0,1,2,\ldots $). Because of the oscillating character of the left-hand
sides of Eqs. (\ref{2.64}), (\ref{2.65}), the system oscillates between the
states with $f_\mu $ and $f_\nu $ with increasing $epHL.$ For larger $epHL$,
the domain of existence of $f_\mu $ becomes narrower, while that of $f_\nu $
widens. For $epHL\rightarrow \infty $, the solution $f_\nu $ goes over
smoothly into that of an infinite multilayer.

We want to point out here that the exact character of the transitions $f_\mu
\leftrightarrow f_\nu $ can only be established by solving the nonlinear
boundary problem and comparing the corresponding free energies, which is
beyond the scope of the present paper.

As an important application of the above results, we consider the critical
temperature of a finite multilayer $T_{cW}$ in the field range given by (\ref
{2.49}), and with $W\ll p\lambda _J/\lambda (T)$. Under such circumstances,
the condition (\ref{2.64}) is satisfied, and the solution of Eq. (\ref{2.59}%
) is
$$
\mu ^2=\frac \alpha {a\xi _0\left( epH\right) ^2}\frac{\Phi _0}{\pi \Phi }%
\left| \sin \frac{\pi \Phi }{\Phi _0}\right| ,
$$
which on substituting into Eq. (\ref{2.60}) yields:

\begin{equation}
\label{2.66}T_{cW}=T_{c0}\left[ 1-\frac{7\zeta (3)}{12}\xi _0^2\chi \left(
\xi _0/l\right) \left[ \frac 13e^2H^2a^2+\frac \alpha {a\xi _0}\left[ 1-
\frac{\Phi _0}{\pi \Phi }\left| \sin \frac{\pi \Phi }{\Phi _0}\right|
\right] \right] \right] .
\end{equation}

Thus, in a finite-size multilayer the critical temperature can exhibit the
same oscillations with changing the flux through the ''elementary cell'' as
the total Josephson current $I$ does [see Eq. (\ref{2.31})]. However, a
significant difference lies in the fact that while the oscillations of $I$
are observable in any types of Josephson systems, the oscillations of $T_c$
is a novel feature, because in a single Josephson junction with thick
superconducting electrodes any shifts of $T_c$ are negligible. In the limit $%
\Phi \gg \Phi _0$, equation (\ref{2.66}) reduces to Eq. (\ref{2.50}), as
anticipated.

\section{DISCUSSION}

Based solely on the microscopic Hamiltonian (\ref{1.1}), we have constructed
a self-consistent theory that provides a comprehensive, unified picture of
physical effects in S/I multilayers in parallel magnetic fields in the GL
regime.

Employing rigorous technique of variational calculus, we have derived in
Sec. II fundamental constraint relations (\ref{1.28}), (\ref{1.29}) and
solved a nontrivial problem of exact minimization of the microscopic
functional (\ref{1.5}). Up until the present study, such a problem has not
been solved even for the much simpler phenomenological LD functional.
Surprisingly, even mutual dependence and incompleteness of the
Euler-Lagrange equations for $\phi _n$ and ${\bf A}$ were not noticed in
previous publications. This incompleteness fully manifests itself in
unphysical degrees of freedom and an irrelevant length scale of equations
for the phase differences proposed, e. g., in Refs. (\cite{BC91,BCG92}):
Making use of constraints of the type (\ref{1.29}), one can reduce these
equations to our Eq. (\ref{1.46}) with $f(y)=1$. Emerging as a direct
mathematical consequence of such general physical properties as gauge
invariance and Josephson interlayer coupling, constraints (\ref{1.28}), (\ref
{1.29}) should apply to any superconducting weakly coupled periodic
structure. The discovery of their role in minimizing the free energy makes
further progress in the development of the theory possible.

In the thin-layer limit which corresponds to the domain of validity of the
phenomenological LD model, we have derived a remarkably simple, closed set
of self-consistent microscopic mean-field equations (\ref{1.43})-(\ref{1.49}%
) and the generating functional (\ref{1.50}). The fact that the solution (%
\ref{2.18}), (\ref{2.21}) and (\ref{2.24}) of these equations describing the
vortex state in an infinite multilayer reproduces the result obtained by
Theodorakis\cite{Th90} in the framework of the LD model is not an occasional
coincidence. The application of the mathematical methods of this paper
allows us to obtain the complete exact solution of the LD model in parallel
fields as well. (This solution will be published elsewhere.) The resulting
mean-field equations are merely a limiting case of our Eqs. (\ref{1.43})-(%
\ref{1.49}) for $a/p\rightarrow 0$. As our equations contain more physical
information, we propose that they should replace the LD model in parallel
fields.

Concerning some major physical results of Sec. III in the thin-layer limit,
we have completely revised previous calculations\cite{B73,CC90} of $H_{c1}$
based on an invalid assumption of single Josephson vortex penetration and
refuted the concept\cite{BC91} of a triangular Josephson vortex lattice. Our
consideration envisages simultaneous and coherent penetration in the form of
the ''vortex planes''. Our prediction of the superheating field $H_s$ for
semi-infinite multilayers implies hysteretic behavior of the magnetization.
In the vortex state, the magnetization should exhibit jumps due to the
vortex-plane penetration. We have fully clarified the wide-spread\cite
{BCG92,FG94} misunderstanding of the physics of the Fraunhofer oscillations:
our self-consistent treatment of the Josephson effect unambiguously proves
that the Fraunhofer pattern occurs due to successive penetration of the
vortex planes and their pinning by the edges of the sample. Our prediction
of novel size-effects in finite multilayers, a series of first-order phase
transitions to the normal state and oscillations of the critical temperature
versus the applied field, should stimulate further experimental
investigation.

The results of our investigation directly apply to artificial
superconductor/insulator\cite{Ne93} and superconductor/semiconductor\cite
{RBB80,V96,YFOF98} multilayers. As regards the high-$T_c$ superconductors
BSCCO and TBCCO, believed to be atomic-scale weakly coupled superlattices,%
\cite{KM94} the application is restricted by the limitation (\ref{1.4}).
However, we expect that such basic features of the thin-layer limit as
simultaneous and coherent penetration in the form of the vortex planes will
hold. For high-$T_c$ samples exhibiting a clear Fraunhofer pattern,\cite{L96}
we anticipate the presence of the related effect of oscillations of the
critical temperature (\ref{2.66}) as well.

As to direct experimental verification of basic concepts of our theory, the
best evidence is provided by the recent magnetization and polarized neutron
reflectivity measurements on Nb/Si multilayers in parallel fields.\cite
{YFOF98} These measurements clearly revealed simultaneous penetration of
Josephson vortices into all Si layers and a companion effect of jumps of the
magnetization, exactly as predicted in our paper. The distribution of the
magnetic field attributed to Josephson vortices was found to have the
periodicity of the Nb/Si layering, in agreement with the general
consideration of Sec.II. A. (The experimental conditions\cite{YFOF98} did
not fully match the requirements of the thin-layer limit for which the
screening by the intralayer currents could be neglected.) Finally, it is
quite natural that our general implicit expression (\ref{2.51}) for $%
H_{c2\infty }(T)$ exhibits the so-called ''3D-2D crossover'', well-known
from the experiment,\cite{RBB80} and is free from the unphysical
''low-temperature'' divergence, typical\cite{T96} of the phenomenological LD
model.

\bigskip

\begin{center}
{\bf ACKNOWLEDGEMENTS}
\end{center}

\smallskip\ The author thanks \v S. Be\v na\v cka, I. I. Fal'ko, V. N.
Morgoon, M. A. Obolenskii, A. N. Omel'yanchuk, I. V\'avra, and S. Tak\'acs
for helpful discussions.

\appendix

\section{The application of Mathieu functions}

The canonical form of the Mathieu equation is\cite{AS,M64}
\begin{equation}
\label{c.1}\frac{d^2f}{dt^2}+\left( a-2q\cos 2t\right) f=0.
\end{equation}

If $f(t)$ is a solution to (\ref{c.1}), then $f\left( \frac \pi 2-t\right) $
satisfies
\begin{equation}
\label{c.2}\frac{d^2f}{dt^2}+\left( a+2q\cos 2t\right) f=0.
\end{equation}

In the class of periodic solutions of (\ref{c.1}), the smallest eigenvalue $%
a_0(q)$ is a non-positive, continuous, even, monotonously decreasing
function of $q$. The corresponding eigenfunction ce$_0(t,q)$ has a period $%
\pi $, is even and strictly positive.

For $0\leq $$q\ll 1$, we have the asymptotics
\begin{equation}
\label{c.3}\text{ce}_0(t,q)\approx \frac 1{\sqrt{2}}\left[ 1-\frac q2\cos
2t+\ldots \right] ,
\end{equation}
\begin{equation}
\label{c.4}a_0(q)\approx -\frac{q^2}2+\ldots
\end{equation}

For $q\gg 1$,
\begin{equation}
\label{c.5}a_0(q)\sim 2\sqrt{q}-2q,
\end{equation}
but there is no uniform asymptotics for ce$_0(t,q).$ In this case, the
behavior of ce$_0(t,q)$ may be characterized by the formulas
\begin{equation}
\label{c.6}\text{ce}_0(t,q)\sim \left( \frac \pi 2\right) ^{1/4}q^{1/8}e^{-
\sqrt{q}\cos {}^2t/4},\quad \left| \cos t\right| <\frac{2^{1/4}}{\sqrt{q}};
\end{equation}
\begin{equation}
\label{c.7}\text{ce}_0(0,q)\sim 2\sqrt{2}e^{-2\sqrt{q}}\text{ce}_0(\pi
/2,q)\sim 2\left( 2\pi \right) ^{1/4}q^{1/8}e^{-2\sqrt{q}}.
\end{equation}

\newpage\

\begin{center}
{\bf FIGURES}
\end{center}

\bigskip\

\begin{description}
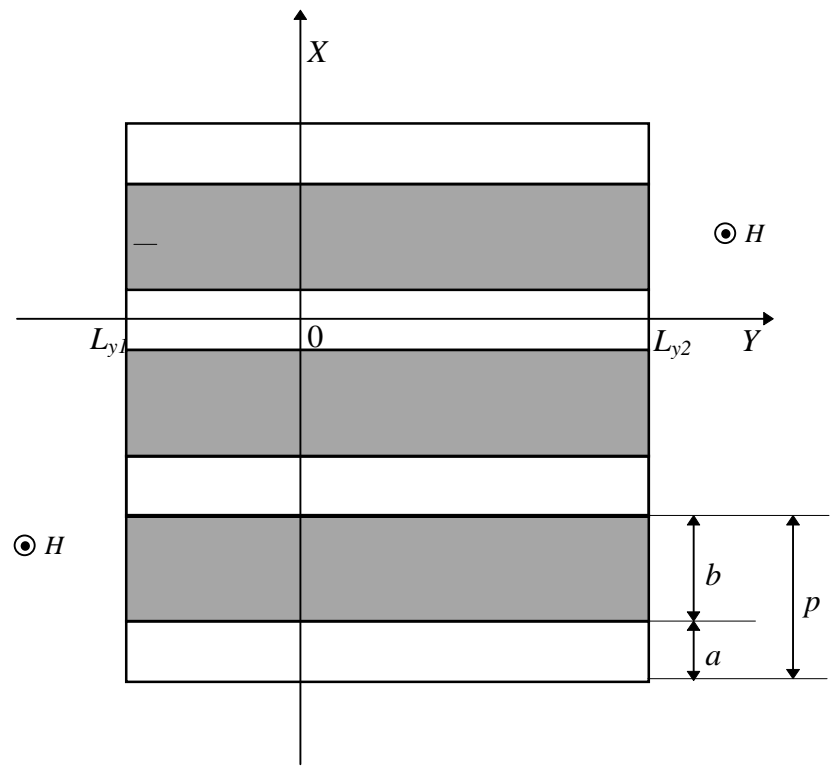
\item  Fig. 1. Geometry of the problem. Alternating superconducting layers
and nonsuperconducting barriers are shown by white and grey rectangles,
respectively. The system is supposed to be infinite in the $x$ and $z$
directions. An external magnetic field $H$ is applied along the $z$ axis.

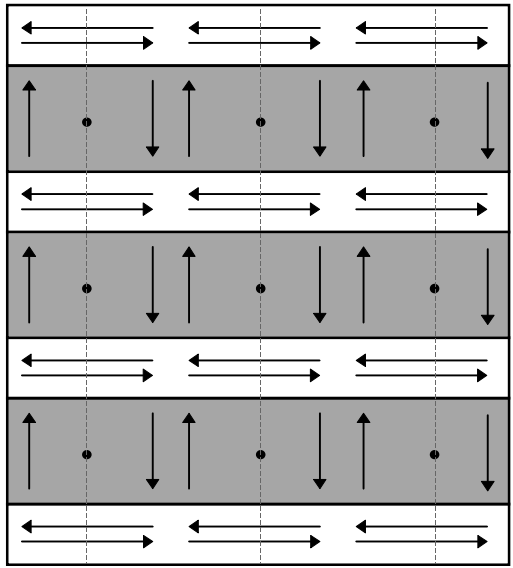
\item  Fig. 2. Vortex state in a finite LD multilayer (in cross section).
Josephson vortices (i. e., singularities of the amplitude of condensation)
are conventionally denoted by black dots. The ''vortex planes'' [i. e.,
maxima of the microscopic magnetic field $h(y)]$ are shown by dashed lines.
Arrows show the direction of supercurrents.
\end{description}

\end{document}